\documentclass[fleqn,usenatbib]{mnras}
\usepackage{newtxtext, newtxmath}
\usepackage[T1]{fontenc}

\DeclareRobustCommand{\VAN}[3]{#2}
\let\VANthebibliography\thebibliography
\def\thebibliography{\DeclareRobustCommand{\VAN}[3]{##3}\VANthebibliography}

\usepackage{graphicx}	% Including figure files
\usepackage{amsmath}	% Advanced maths commands

\title[NGC\,628 Bubble Ages]{Ages of the resolved stellar populations inside the JWST/MIRI bubbles in NGC\,628}

\author[A. Ck. et al.]{
Avinash Ck$^{1}$,
Divakara Mayya$^{1}$,
Alessandro  Bressan$^{2,3,4}$,
Jairo Andres Alzate Trujillo$^{5}$,
Léo Girardi$^{3}$, and
\newauthor
Bolivia Cuevas-Otahola$^{6}$\thanks{E-mail: b.cuevas.otahola@gmail.com}
\\
$^{1}$Instituto Nacional de Astrofísica, Óptica y Electrónica, Luis Enrique Erro 1, Tonantzintla 72840, Puebla, México\\
$^{2}$SISSA, Via Bonomea 265, I-34136 Trieste, Italy\\
$^{3}$Osservatorio Astronomico di Padova—INAF, Vicolo dell’Osservatorio 5, I-35122 Padova, Italy\\
$^{4}$Purple Mountain Observatory, Chinese Academy of Sciences, Nanjing 210023, People’s Republic of China\\
$^{5}$Instituto de Astrofísica de Andalucía – Consejo Superior de Investigaciones Científicas (IAA-CSIC), 
S/N,E-18008, Granada, Spain\\
$^{6}$Instituto de Astronomía, Universidad Nacional Autónoma de México, Box 70-264, México City, México}
\date{Accepted XXX. Received YYY; in original form ZZZ}

\pubyear{\the\year{}}

\begin{document}
\label{firstpage}
\pagerange{\pageref{firstpage}--\pageref{lastpage}}
\maketitle

\begin{abstract}
JWST images in the MIRI filters are characterized by prominent interstellar bubbles, most of which are expected to be created by mechanical energy injected into the interstellar medium by dying massive stars. In this work, we use resolved stellar populations (RSPs) in JWST/NIRCam images of NGC\,628 from the JWST-FEAST dataset to determine the demography of stellar populations within these bubbles by comparing them with PARSEC+COLIBRI isochrones using a Bayesian framework. Our analysis reveals the presence of multiple stellar populations younger than 100~Myr within the bubbles, suggesting secondary star formation following the first generation of stars that formed the bubble. We find a clear preference for the most recently formed stars to lie closer to the bubble shell. In contrast, relatively older stars are distributed throughout the bubble interior, consistent with a scenario in which stars formed in dense shells are left behind as the shell expands. We also examine PHANGS-HST cluster candidates within large bubbles and find no convincing progenitor clusters responsible for the initial trigger, indicating that low-mass clusters or OB associations may be sufficient to drive the initial expansion. However, the cluster may dissolve as it co-evolves with the bubble, producing the dispersed stellar distributions observed in larger bubbles. We establish a fundamental plane relation for stellar feedback-driven bubbles that involves the stellar population mass, age, bubble size, gas density, and feedback efficiency, highlighting the ability of JWST/NIRCam CMDs to characterize stellar populations driving interstellar bubble expansion in nearby galaxies.

\end{abstract}


%%%%%%%%%%%%%%%%%%%%%%%%%%%%%%%%%%%%%%%%%%%%%%%%%%

%%%%%%%%%%%%%%%%% BODY OF PAPER %%%%%%%%%%%%%%%%%%

\section{Introduction}
Bubbles are voids in the interstellar medium (ISM) of galaxies formed by feedback from massive stars \citep[e.g.][]{stellar_feedback, v_exp1, bubble_intro1, bubble_intro2, bubble_intro3, bubble_intro4, bubble_intro5}. Stellar feedback processes such as stellar winds and supernovae inject energy and momentum into the ISM, producing expanding shells and filaments \citep{tenerio_tagle}. As feedback accumulates over time, expanding shells become superbubbles with radii of hundreds of parsecs \citep{superbubble_mult_gen}. Observations from the James Webb Space Telescope's (JWST) Mid-Infrared Instrument (MIRI), particularly from surveys like JWST-PHANGS \citep{phangs}, have revealed a completely new way to detect superbubbles. This is because most of the MIRI filters (e.g., F770W) intercept spectral features from polycyclic aromatic hydrocarbons (PAHs) that originate from the diffuse interstellar medium as well as photo-dissociation regions (PDRs) around H\,{\sc ii} regions. Bubbles show up as bright-rimmed, almost circular structures as the energetic UV photons from stellar feedback processes destroy the PAHs inside them \citep[e.g.][and references therein]{pah_dest} and the stellar feedback sweeps up material, forming bright PAH shells with high flux contrast \citep [e.g.][]{bubble_shell}.
One of the most striking examples among JWST-PHANGS galaxies is the late-type spiral NGC\,628 (M74), which hosts a kiloparsec-scale superbubble known as the "Phantom Void" \citep{Barnes_2023}. 

Bubble formation around massive stars had been theoretically predicted almost half a century ago \citep{castor,stellar_feedback}. \citet{tenerio_tagle} discussed the effect of stellar feedback on large scales in galaxies, including the creation of large-scale outflows seen in starburst galaxies such as M\,82 \citep{m82_outflow}. Before JWST observations, bubbles were traced mainly as HI holes using the 21-cm radio emission \citep{hi_hole1,hi_hole2}), or in ionized gas (e.g., H$\alpha$ emission) in galaxies \citep{halpha_hole1,halpha_hole2}. The use of HI holes as a bubble detection strategy was limited by the relatively low shell-to-bubble contrast and the typical few arcsec spatial resolution available, even with the interferometric arrays, whereas the ionized shells are often broken or incomplete, making it hard to trace all existing shells in a galaxy. The subarcsec spatial resolution, the high shell-to-bubble contrast of the PAH emission, and a field of view large enough to cover the star-forming disks in the MIRI/F770W images have allowed the possibility of detecting all bubbles larger than a few parsecs in nearby star-forming galaxies. The availability of a uniform catalog of bubbles opens up an opportunity to address the role of stellar feedback in disk galaxies with moderate star formation. We aim to address this issue here in NGC\,628, the most iconic of the JWST-PHANGS galaxies \citep{phangs}. 

A comprehensive study of stellar feedback processes requires knowledge of the stellar populations responsible for creating the bubbles \citep[e.g.][]{bubble_hst_cmd}. The Hubble Space Telescope (HST) and the JWST/NIRCam have filters that trace direct emission from stellar populations of diverse ages, thus allowing a comprehensive study of the formation, energetics, and survival of the bubbles in normal disk galaxies. 

\citet{superbubble} used HST color-magnitude diagrams (CMDs) to construct the star formation history (SFH) of stars inside the Phantom Void, finding that it formed from feedback by stellar populations over the past $\sim$50 Myr. The study also revealed spatially segregated, sequential star formation, indicating star formation in the expanding bubble. 

In this work, we extend the resolved stellar population studies to more bubbles in NGC\,628 to throw light on the typical time-scale of survival of JWST/MIRI bubbles. In particular, we investigate the age and star-formation history of stellar populations residing inside the bubbles using the CMDs constructed with the JWST/NIRCAM images in  F115W, F150W, and F200W filters, available
from the JWST-FEAST\footnote{Feedback in Emerging extrAgalactic Star clusTers} survey \citep{feast}.
NGC\,628 is a mildly star-forming galaxy with a star formation rate of $\sim$1.8~M$_{\sun}$/yr ~\citep{NGC628_muse_SFR} and is nearly face-on with an inclination of i=$9^{\circ}$ \citep{inc_1, inc_2, NGC628_sfr2} and a position angle of the major axis of 21$^{\circ}$ \citep{NGC628_sfr2}. The galaxy has been a target of several studies ranging from metallicities, gas kinematics, HII regions, and star formation \citep{NGC628_sfr1, NGC628_sfr2, NGC628_muse_SFR}. In a recent study, \cite{trgb_NGC628} revised its distance from the usually used distance of 9.8~Mpc \citep{hst_trgb_NGC628} to 9.16~Mpc based on the JWST-based tip of the red giant branch (TRGB) measurements, which we adopt in this work. Section 2 describes the bubble sample and stellar photometry. Section 3 describes the methodology utilized for obtaining ages of stars and SFHs from CMDs. In Section 4, we present properties of stellar populations within the bubbles derived from CMD analyses and how they are connected to the energetics of the feedback.

\begin{figure*}
    \centering
    \includegraphics[width=\linewidth]{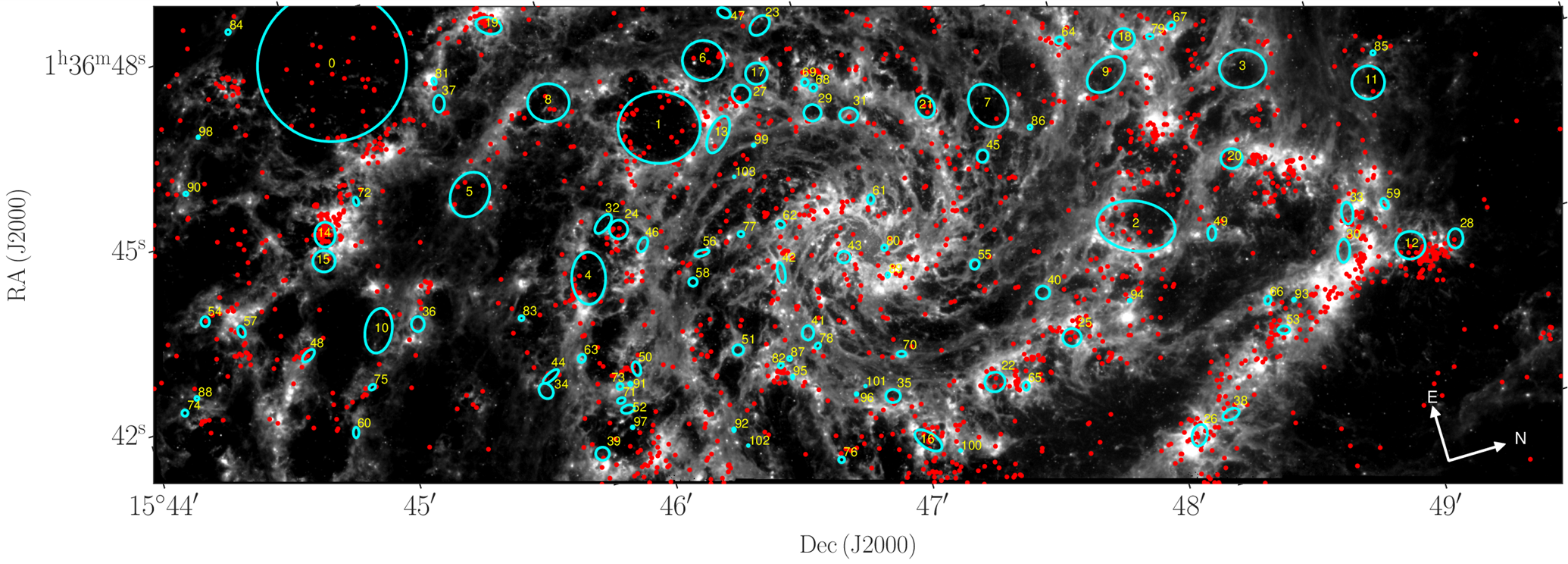}
    \caption{JWST/MIRI F770W image of NGC\,628 covering a Field of View of $6\arcmin \times 2\arcmin$ field. The cyan ellipses indicate bubbles, along with their identification numbers. The red-filled circles indicate Supergiants (SGs) selected using the CMD in NIRCam filters F115W and F200W. The largest of the bubbles (bubble \#0 at the top-left) has a radius of 815~pc. The North-East compass is shown at the bottom-right corner to guide the orientation of the RA-DEC tick marks.}
    \label{fig:bubble_sample}
\end{figure*}

\section{The bubble sample and stellar catalog}

\subsection{Bubble sample}

\begin{figure}
    \centering
    \includegraphics[width=\columnwidth]{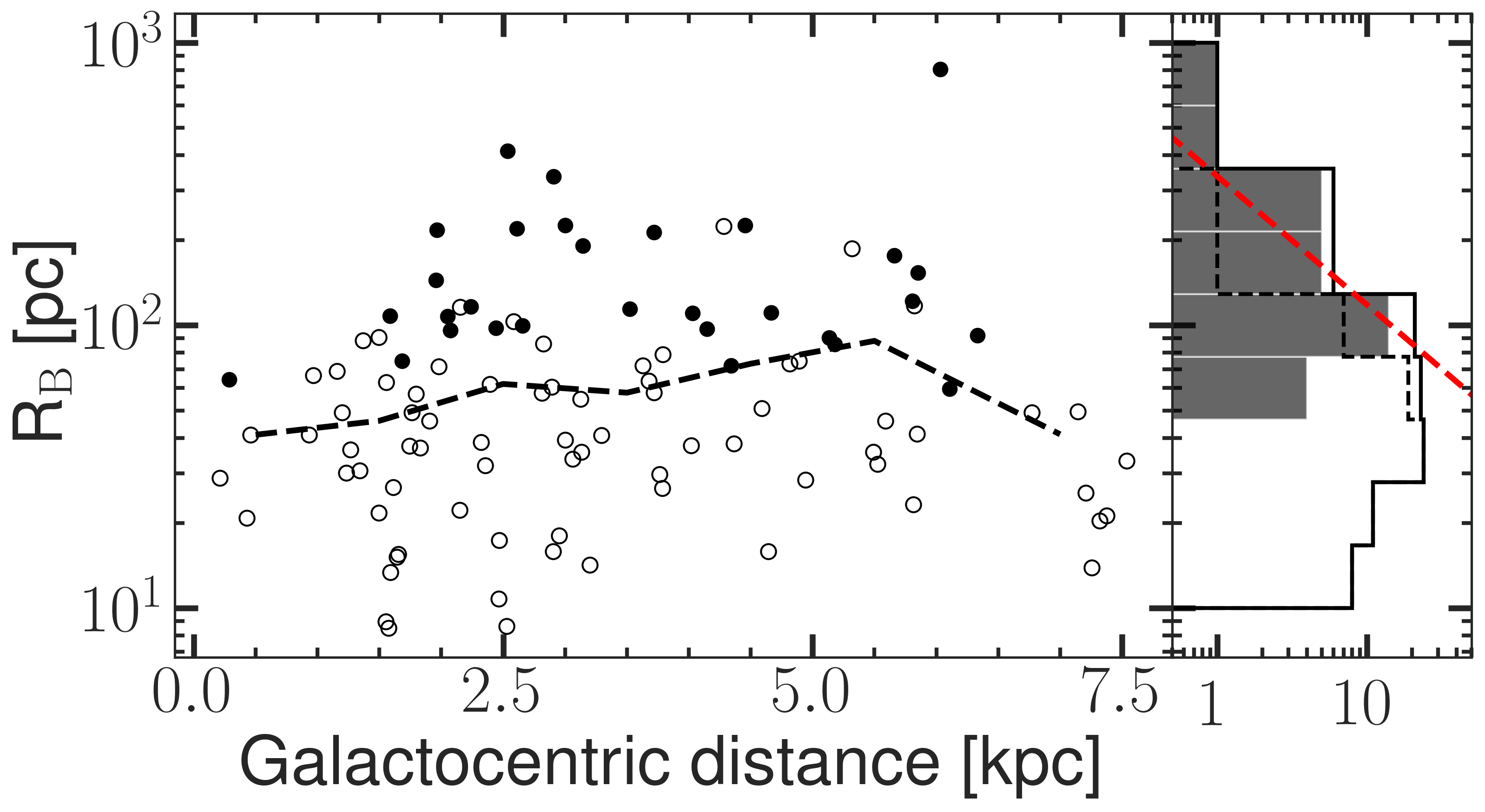}
    \caption{\textit{Left:} Bubble radius vs galactocentric distance. Bubbles with at least 10 massive stars (filled circles) are distinguished from the rest. The median bubble radius  (dashed line) shows a tendency to increase from 40~pc near the centre to 80~pc at 5.5~kpc, decreasing again to 40~pc at 7~kpc. \textit{Right:} Histogram of bubble radii (line). Bubbles containing at least 10 massive stars are indicated by a solid histogram. The dashed step plot shows bubbles containing fewer than 10 massive stars. The red dashed line shows the distribution for the W23 sample, which is a power law with an index of -2.2.}
    \label{fig:bubble_dist}
\end{figure}
In the present work, we aim to study a representative sample of bubbles spanning a diverse range of sizes. In the massive star feedback-driven scenario of bubble formation, the sizes of bubbles are set by the density of the gas and the cumulative energy output from the massive stars \citep[e.g.][]{bubble_dist1, nath_2020, phantom_void_sim}.  The energy and momentum output from massive stars make the bubbles expand over time. The bubbles grow larger at lower ISM densities and with higher energy output from massive stars. The energy output originates from stellar winds at ages less than $\sim$5~Myr, with the Type II supernovae (SNe) contributing to it at older ages up to around 40~Myr when the last of the massive stars explodes as SNe \citep{starburst99}. Given the prolonged duration of SNe explosions, it is expected that the larger bubbles are systematically older than the smaller bubbles and vice versa, assuming ISM density does not vary much from region to region. A study of the ages of stellar populations responsible for bubbles of a diverse range of sizes would allow us to test these theoretical expectations, which were mostly carried out for idealized conditions.

While bubble-shaped structures are easily perceived by the human eye in the JWST/MIRI images, counting the bubbles is a complex problem, as bubbles exhibit a wide range of sizes, shapes, and flux contrasts. Often, the boundary walls (referred to as shells hereafter) of the bubbles are discontinuous and non-uniformly illuminated. Furthermore, it is common to have smaller bubbles nested near the shells of the larger bubbles. All these problems make an automated approach highly incomplete as of now. \citet{Watkins_2023} carried out a first census of bubbles on a JWST image, visually counting more than 1000 bubbles in NGC\,628. They found that the manually fitted sizes obey a power-law distribution with an index of $-2.2$ spanning a bubble radius between 20 and 800~pc. Their observed slope is close to that expected for star-forming galaxies in hydrodynamical simulations of \cite{nath_2020}. To guard against the results being affected by the subjectivity of the human eye, they presented the results for three samples independently obtained by three of the co-authors, finding that the statistical results obtained for the three samples are identical within the errors, although the number of bubbles and their exact definition varied among the three samples. They found that the large bubbles often contain a number of smaller bubbles within them.  

For the purpose of our study here, we require a representative bubble sample covering the entire range of sizes rather than a complete census of bubbles. Given this objective, we visually identified a sample of stellar-feedback-driven bubble candidates in the NGC\,628 using the JWST-MIRI F770W image, following a strategy similar to that of \citet{Watkins_2023}.  A bubble is characterized by a low flux followed by a sharp increase in the PAH emission traced by the F770W image. This sharp increase is because of the accumulation of the swept-up material forming the shell structure. To facilitate a neat association of bubbles with the stellar population responsible for their creation, we ensured that the selected bubble is well clear of any other bubble of comparable or larger size. This criterion reduces confusion from overlapping structures without losing the statistical properties of the sample by \citet{Watkins_2023} (henceforth, the W23 sample).

Given that it is the nature of bubble growth that a large bubble contains a lot of smaller bubbles in its periphery, we ignored any such small bubbles as well. On the other hand, we selected smaller bubbles that are isolated from any other bubbles. This resulted in a sample of 104 bubbles, which are shown superposed on the F770W image in Fig.~\ref{fig:bubble_sample}. It can be seen that the selected bubbles are distributed over the entire face of the galaxy, covering a variety of sizes. 

For bubble identification, we used the more recent dataset of NGC\,628 obtained from the FEAST dataset rather than the JWST-PHANGS dataset on which W23 carried out the bubble search, because it covers a slightly wider radial extent. In addition, the JWST-FEAST dataset includes JWST/NIRCAM images covering the same spatial extent as the MIRI images in three broad-band short-wavelength filters, providing stellar data at spatial resolutions around a factor of two better than those possible with the HST images.

To quantify bubble morphology, we inscribe ellipses that fit the inner edge of the bubble shell, avoiding the shell as shown in Fig.~\ref{fig:bubble_sample}.
The bubbles cover an effective radius, defined as $R_{\rm B}=\sqrt{ab}$ ($a$ and $b$, the semi-major and semi-minor axes, respectively), from 8.5~pc to 815~pc. In Table~\ref{tab:bubble_sample}, we present the physical properties of the sample bubbles. The bubbles are numbered from 0 to 103, arranged in decreasing order of $R_{\rm B}$.

The distribution of bubble effective radii as a function of galactocentric distance is shown in Fig.~\ref{fig:bubble_dist}~(left). A trend of median size systematically increasing with galactocentric distance up to 5.5~kpc can be noticed.
In the right panel, we show the size distribution of the bubbles, along with the power law fit from W23. It can be inferred that our sample includes the whole range of bubble sizes in the W23 sample. For the large bubbles ($>$80~pc), our sample includes almost all bubbles from that sample.

The largest bubble (Bubble \#0) with an $R_{\rm B}$=815~pc is a clear outlier in Fig.~\ref{fig:bubble_dist}. The PAH shell of this bubble covers less than 50\% of its circumference. It is the only bubble in the sample whose selection is biased from other datasets, the  HI map \citep{HI_THINGS} in this case. This corresponds to the largest of the HI holes in the large FoV of the HI map. We included this bubble to allow us to shed light on the differences, if any, in the stellar populations inside HI holes and classical stellar feedback-driven bubbles.

\begin{table}
\setlength{\tabcolsep}{3pt}
    \centering
    \caption{Geometrical parameters of our sample of Bubbles$^{\dagger}$. }
    \label{tab:bubble_sample}
     \begin{tabular}{c|c|c|c|c|c|c|c}
     \hline
     ID & RA (J2000) & Dec (J2000) & a & q & P.A. & R$_{\rm B}$  &D$_{\mathrm GC}$\\
        &  hhmmss       & ddmmss         & arcsec &  & deg & pc & kpc\\
     1 &   2   & 3 & 4 & 5 & 6 & 7 & 8\\
    \hline
0 & 01h36m47.14s & +15d45m08.46s &  18.53 & 0.98 & 60 & 815 & 6.03 \\
1 & 01h36m44.57s & +15d46m20.73s &  10.04 & 0.88 & -17 & 418 & 2.53 \\
2 & 01h36m40.68s & +15d48m05.50s &  9.63 & 0.63 & -25 & 340 & 2.91 \\
3 & 01h36m42.72s & +15d48m41.37s &  5.71 & 0.81 & -17 & 229 & 4.45 \\
4 & 01h36m42.47s & +15d45m53.79s &  6.39 & 0.65 & 75 & 228 & 3.0 \\
5 & 01h36m44.39s & +15d45m31.80s &  5.71 & 0.80 & 45 & 227 & 4.28 \\
6 & 01h36m45.44s & +15d46m35.70s &  5.10 & 0.96 & -17 & 222 & 2.61 \\
7 & 01h36m43.35s & +15d47m39.24s &  5.74 & 0.74 & -70 & 219 & 1.97 \\
8 & 01h36m45.51s & +15d45m56.53s &  5.09 & 0.91 & -17 & 215 & 3.72 \\
9 & 01h36m43.29s & +15d48m08.99s &  5.32 & 0.67 & 25 & 193 & 3.14 \\
\hline
\end{tabular}
\parbox{\columnwidth}{\vspace{2pt}\small
    $^{\dagger}$ The largest 10 bubbles are shown here. A complete table with all 104 bubbles is provided in the supplementary material.
    \textbf{Notes to Columns:}~(1)~Bubble ID. (2)~Right Ascension of the bubble center. (3)~Declination of the bubble center. (4)~Semi-major axis of the ellipse inscribing the bubble in arcsec. (5)~Axis ratio of the ellipse. (6)~Position angle of the ellipse with respect to the sky plane, measured counter-clockwise from West. (7)~Effective radius of the bubble in parsec units. (8)~Galactocentric distance of the bubble centre from the centre of the host galaxy in kpc units.}
\end{table}

\subsection{Catalog of Resolved Stellar Populations}

\begin{figure}
    \centering
    \includegraphics[width=\columnwidth]{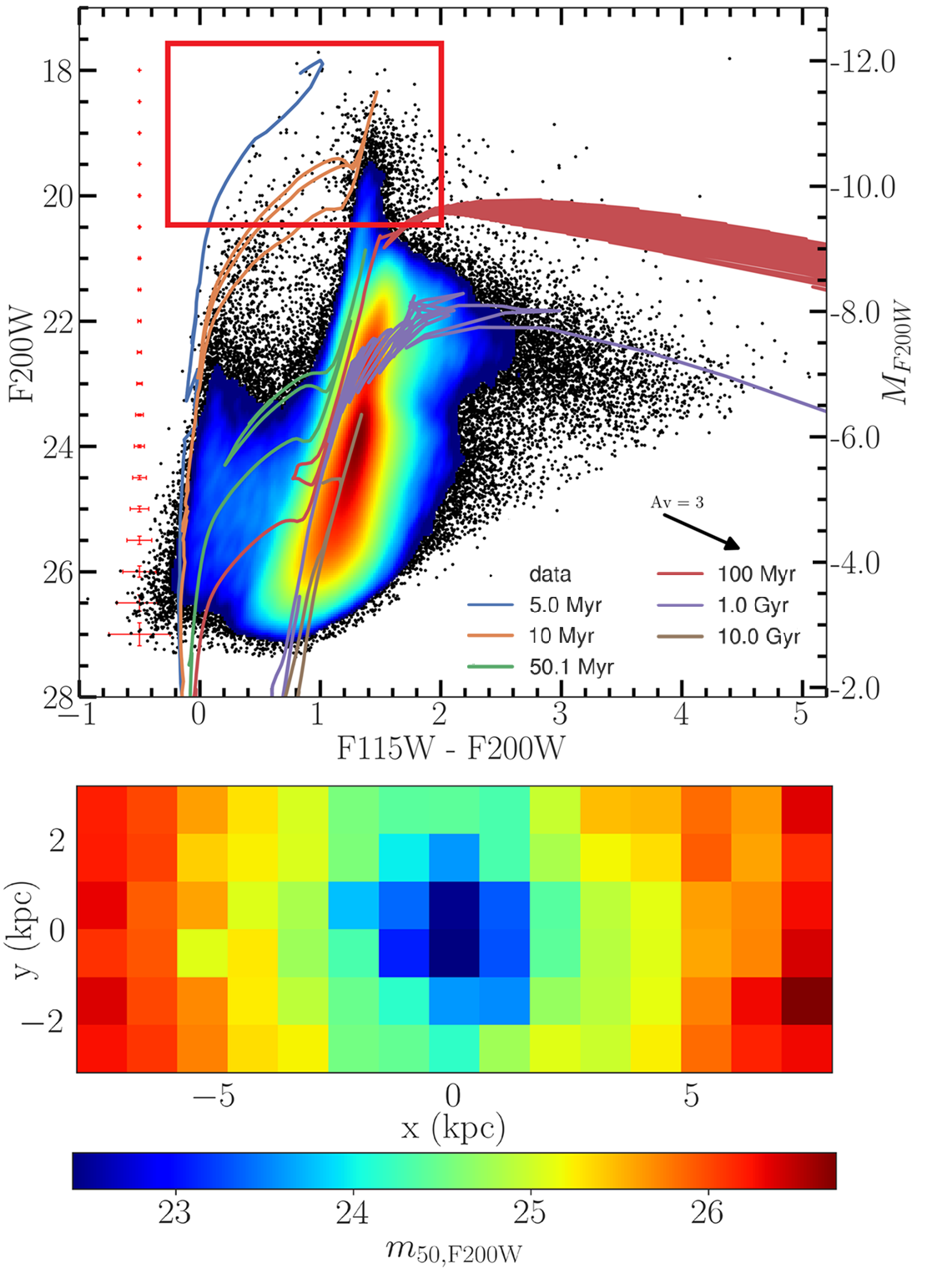}
    \caption{\textit{Top:} NIRCam F200W vs F115W$-$F200W Hess diagram over the entire galaxy disk of NGC 628. The PARSEC+COLIBRI isochrones of selected ages are superposed, with the younger (log(age [yr])$<$9) isochrones being metal-rich (Z = 0.02) than the older ones (Z = 0.002). The red box shows the location dominated by Red and Blue Super Giants (RSGs and BSGs). Vertically aligned red bars show typical error bars on magnitude and color for the complete range of observed magnitudes. \textit{Bottom:} Spatially resolved map of F200W 50\% completeness magnitudes (m$_{50}$) in 24 arcsec$\times$24 arcsec ($\sim$1~kpc$^2$) grids covering the entire field of view of the FEAST image. }
    \label{fig:cmd}
\end{figure}

In a recent study, \citet{trgb_NGC628} obtained photometry of resolved stellar populations in NGC\,628 on JWST-FEAST images in the NIRCam SW filters F115W, F150W, and F200W, using the DOLPHOT routine \citep{dolphot_1,dolphot_2}. We used the photometric catalog and the spatially resolved completeness curves obtained in that work to trace the resolved stellar populations inside the bubbles in our sample. The resulting color--magnitude diagram (CMD) using the F200W and F115W data is shown in Fig.~\ref{fig:cmd}~(Top), displaying approximately 1.1 million stars across a $6\arcmin \times 2\arcmin$ field that spans the NGC\,628 disk. In the same figure, we overlay PARSEC+COLIBRI\footnote{We utilized PARSEC 1.2S + COLIBRI S\_37 + COLIBRI S\_35 + COLIBRI PR16, generated using web interface \href{https://stev.oapd.inaf.it/cgi-bin/cmd\_3.9}{CMD 3.9} with  n$_{\rm inTPC}$=10 and $\eta_{\rm Reimers}$=0.2} \citep{parsec,parsec_colibri, parsec2} isochrones at selected ages and metallicities, adopting a distance modulus of 29.81~mag \citep{trgb_NGC628}, a foreground extinction of $A_V = 0.19$~mag \citep{foreground_av}, and the extinction law from \citet{extinction_law}.

For obtaining spatially resolved completeness curves, we divide the NIRCam field into 24~arcsec$\times$24~arcsec ($\sim$1~kpc$^2$) regions, forming a 15$\times$6 spatial grid. For each grid cell, we perform DOLPHOT artificial star tests across the filters and model the completeness fraction using the Pritchet function \citep{pritchet_1,pritchet_2,pritchet_3,pritchet_4}. The bottom panel of Fig.~\ref{fig:cmd} shows the variation in the 50~per~cent completeness magnitude (m$_{50}$) in the F200W filter across the grid.  

We obtained a catalogue of stars located within the ellipses defining each bubble. The availability of the age of each star in the catalog helps us to classify the bubbles based on the number of massive stars (age$<$50~Myr). 
For the majority of the bubbles, the number of massive stars was insufficient to construct a statistically significant SFH, whereas the largest bubbles contained more than 2000 stars (See N$_{\mathrm{stars}}$ column in Table~\ref{tab:bubble_properties}). Using simulated CMDs, we found that at least 10 young stars (age$<$50~Myr\footnote{The stellar ages are derived using the resolved stellar population method described in Section~\ref{sec:baye_age}}) are required to recover a reliable SFH associated with bubble formation. Bubbles containing at least 10 massive stars are systematically bigger than those containing fewer than 10 massive stars. Applying this criterion yields a sub-sample of 30 large bubbles (R$_{\mathrm{B}}>60$~pc, henceforth, big bubble sample) and 74 small bubbles. The right panel of Fig.~\ref{fig:bubble_dist} compares the size distribution of the big and small bubble samples against the overall bubble size distribution of bubbles.

\section{Analysis of the CMDs to determine the SFHs of the bubble stars}
\label{sec:3}

The CMD of a region carries information on the ages of all the stars belonging to the region. A region like a bubble contains at least two distinct populations of stars: (1) stars responsible for the creation of the bubble, and (2) stars belonging to the underlying disk. Here, by underlying disk stars, we refer to stars that are not associated with the star formation related to the bubble.
These stars are detected within the bubble region purely due to projection effects. The position of the stars in the CMDs is dictated by the metallicity and interstellar extinction, both of which could be different for the two populations. We describe in this section the procedure we have followed to take into account these differences. 

\subsection{Interstellar extinction and age-dependent metallicities of bubble stars}
\begin{figure}
    \centering
    \includegraphics[width=\columnwidth]{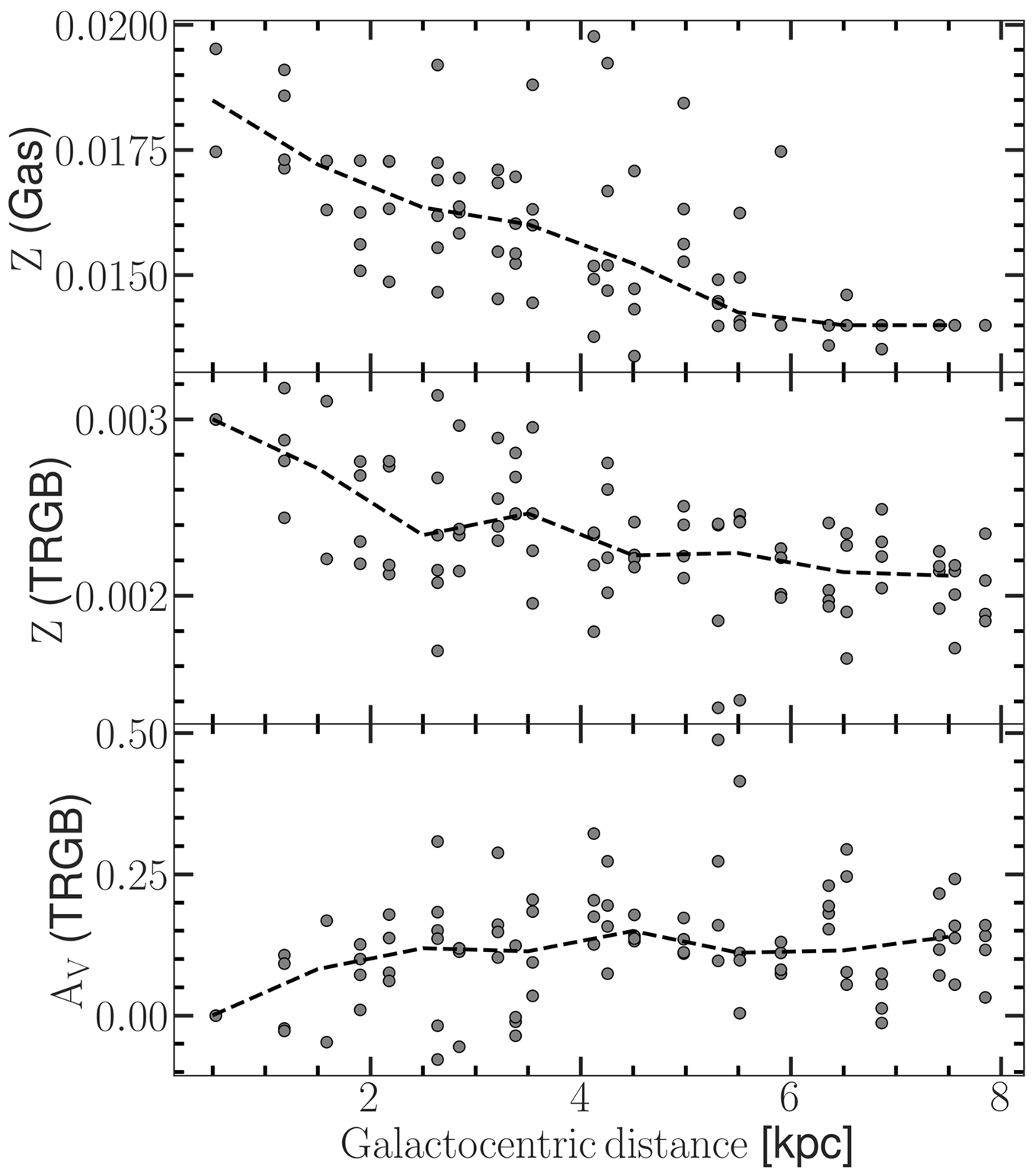}
    \caption{Metallicities and diffuse ISM extinctions in $\sim$1~kpc$^2$ square regions, spanning the NIRCam FoV. \textit{Top:} Gas-Phase metallicity derived using N2 relation with MUSE data. \textit{Middle:} Metallicity obtained from TRGB color-magnitude measurements. \textit{Bottom:} A$_V$ of the diffuse ISM obtained from TRGB color-magnitude measurements. The dashed line joins the median values in radial bins of 1~kpc width (see Sec.3.1 for details).}
    \label{fig:met}
\end{figure}

A key challenge in recovering ages of resolved stellar populations using the isochrones is the degeneracy among age, metallicity, and extinction. For young populations, metallicity can be constrained by the nebular metallicities derived for the ionized gas. PHANGS-MUSE dataset for NGC\,628 provides such information \citep{phangs_muse}. We used this dataset and the N2 \cite{N2} relation to obtain mean current metallicities. We have used a solar 12+log(O/H)=8.69 \citep{n2_sol} and Z=0.017 \citep{n2_sol_z} for converting our nebular abundances to metallicities in grids of 24 arcsec$\times$24 arcsec over the entire image. The resulting metallicity in each grid is shown as a function of the galactocentric distance of the grid centre in the top panel of Fig.~\ref{fig:met}. 

\citet{trgb_NGC628} obtained the metallicity of TRGB stars in each of the 90 grids. These metallicities are representative of old stellar systems in the disk. The TRGB metallicity for each grid is shown in the middle panel of Fig.~\ref{fig:met}. Like the nebular metallicity, the old stellar metallicity also shows a gradient, but the values in general are a factor of eight smaller.

The interstellar extinction from the PHANGS-MUSE nebular dataset is also available for NGC\,628. However, these extinction values correspond to those in star-forming regions whose extinction properties are distinct from the relatively gas-free regions inside the bubble. The empty nature of the bubble ensures that the stellar populations inside the bubble experience little extinction. \citet{trgb_NGC628} obtained extinction in diffuse gas in this galaxy using the TRGBs, which are plotted in the bottom panel of Fig.~\ref{fig:met}.  In general, $A_V$ values lie below 0.35~mag, with a median value of 0.12~mag.

\subsection{Bayesian method of  determination of age and SFH}

\begin{table}
 \setlength{\tabcolsep}{2.1pt}
    \centering
    \caption{Derived stellar population properties of bubbles.}
    \label{tab:bubble_properties}
     \begin{tabular}{c|c|c|c|c|c|c|c|c|c}
     \hline
     ID & Grid ID & t$_{\xi}$ &  m$_{\rm 50,F200W}$ & $t_0$ & $t_{25}$ & $t_{50}$ & $t_{75}$ & $M_{*}$ & $N_{\mathrm{stars}}$\\
        &   & Myr & mag & Myr   &  Myr     & Myr      & Myr      & log M$_{\sun}$ & \\
        1 & 2 & 3 & 4 & 5 & 6 & 7 & 8 & 9 & 10\\
    \hline
        0 &  80 & 100.0 & 25.9 & 100.0 & 86.4 & 66.0 & 50.4 & 6.5 & 12217 \\
    1 &  56 & 50.1 & 24.6 & 39.8 & 37.5 & 31.7 & 21.5 & 5.8 & 5744 \\
    2 &  26 & 50.1 & 24.6 & 39.8 & 37.8 & 31.9 & 15.4 & 5.3 & 3375 \\
    3 &  18 & 63.1 & 25.2 & 39.8 & 37.9 & 29.7 & 13.6 & 4.8 & 1269 \\
    4 &  63 & 50.1 & 24.9 & 50.1 & 48.4 & 40.5 & 26.4 & 5.4 & 1565 \\
    6 &  55 & 50.1 & 24.6 & 25.1 & 17.3 & 14.2 & 12.6 & 4.8 & 1398 \\
    7 &  37 & 31.6 & 24.2 & 25.1 & 20.2 & 11.7 & 7.5 & 4.6 & 1390 \\
    8 &  62 & 50.1 & 24.9 & 31.6 & 22.8 & 13.5 & 7.5 & 4.8 & 1339 \\
    9 &  30 & 50.1 & 24.6 & 31.6 & 18.9 & 9.5 & 7.3 & 4.5 & 1115 \\
    11 &  12 & 100.0 & 25.7 & 39.8 & 36.3 & 30.9 & 23.2 & 5.0 & 631 \\
    12 &  8 & 63.1 & 25.5 & 39.8 & 25.0 & 18.3 & 11.9 & 5.4 & 464 \\
    13 &  56 & 31.6 & 24.1 & 31.6 & 23.5 & 12.1 & 7.7 & 4.6 & 622 \\
    14 &  75 & 100.0 & 25.7 & 39.8 & 28.2 & 18.3 & 13.5 & 5.6 & 339 \\
    16 &  40 & 31.6 & 24.2 & 31.6 & 26.7 & 16.9 & 9.1 & 5.2 & 441 \\
    18 &  30 & 50.1 & 24.7 & 25.1 & 15.5 & 12.4 & 8.4 & 4.5 & 359 \\
    19 &  67 & 63.1 & 25.2 & 63.1 & 43.8 & 20.2 & 11.7 & 5.5 & 329 \\
    20 &  19 & 63.1 & 25.1 & 63.1 & 60.0 & 40.4 & 26.8 & 5.2 & 387 \\
    21 &  43 & 25.1 & 23.8 & 25.1 & 23.2 & 15.1 & 8.2 & 4.7 & 306 \\
    22 &  34 & 31.6 & 24.0 & 31.6 & 24.4 & 18.0 & 11.1 & 4.9 & 331 \\
    24 &  57 & 50.1 & 24.6 & 39.8 & 29.1 & 20.4 & 13.3 & 4.9 & 320 \\
    25 &  33 & 31.6 & 24.1 & 31.6 & 29.4 & 15.7 & 10.6 & 5.0 & 232 \\
    26 &  22 & 50.1 & 25.2 & 50.1 & 45.5 & 30.1 & 14.2 & 5.2 & 249 \\
    27 &  49 & 31.6 & 24.0 & 31.6 & 26.1 & 13.4 & 10.0 & 4.8 & 258 \\
    28 &  8 & 125.9 & 25.9 & 63.1 & 35.1 & 21.0 & 11.9 & 4.8 & 135 \\
    30 &  14 & 63.1 & 25.1 & 39.8 & 31.7 & 12.3 & 10.5 & 4.7 & 198 \\
    33 &  13 & 63.1 & 25.4 & 50.1 & 36.6 & 24.5 & 15.2 & 5.0 & 166 \\
    35 &  40 & 25.1 & 23.7 & 25.1 & 23.6 & 17.8 & 9.6 & 4.8 & 157 \\
    38 &  22 & 63.1 & 25.2 & 63.1 & 55.6 & 39.8 & 22.1 & 4.9 & 142 \\
    43 &  45 & 25.1 & 22.4 & 25.1 & 24.3 & 16.2 & 7.7 & 4.9 & 121 \\
    48 &  76 & 100.0 & 25.7 & 50.1 & 35.8 & 20.1 & 12.8 & 4.7 & 63 \\
\hline
\end{tabular}
\parbox{\columnwidth}{\vspace{2pt}\small
    \textbf{Notes.}~Columns: (1)~Bubble ID. (2)~ID of the grid in which the bubble is located. (3)~Youngest age bin at which incompleteness correction is greater than 1.5. (4)~F200W m$_{50}$. (5--8)~Derived quantities from the SFH. (9)~Integrated stellar mass of stellar populations inside the bubble. (10)~Number of stars inside the bubble above completeness limits.}
\end{table}
\label{sec:baye_age}
The SFH of the populations inside the bubbles is obtained employing the statistical model from \citep{bayesfh, superbubble}. This method applies Bayes' theorem in a hierarchical mode to infer the SFH of any resolved stellar population based on the photometry of its stars. Within this framework, the algorithm determines the probability distribution of stars' positions in the CMD by expressing them as a linear combination of theoretical isochrones of different ages. Each isochrone is assigned a weight $a_i$, which corresponds to the fraction of the total stellar population that aligns with that particular isochrone \citep{Small2013}. 
This statistical approach differs from the traditional method of retrieving the SFH by minimizing the number of stars in bins of colors and magnitudes in a CMD \citep[e.g., MATCH,][]{match}. Our approach enables us to utilize existing data across all available filters. Additionally, this approach also provides a best-fit age for each star. More details about the Bayesian model are shown in Appendix \ref{sec:bay_model}.

In the present work, we utilized JWST/NIRCam F115W, F150W, and F200W photometry and completeness estimates to determine the ages of stellar populations inside bubbles. We use 34 stellar isochrones from PARSEC+COLIBRI between 6.7$\leq\log$(Age\,[yr])$\leq$10 with $\Delta\log$(Age\,[yr])=0.1~dex. The stellar masses between 0.1 and 350 $M_{\sun}$ are used in the isochrones, along with \citet{imf} initial mass function (IMF). The isochrones of all ages are reddened by the $A_V$ value corresponding to the grid under analysis. The metallicities of the isochrones are chosen to be the nebular metallicity nearest to $Z \in \{0.014, 0.017, 0.02\}$ in the grid where a given star is located if the isochrones' ages are younger than 1~Gyr, and that of the TRGBs~($Z \in \{0.001, 0.002, 0.003\}$) if they are older. The choice of 1 Gyr as the delimiting age for metal-rich and metal-poor systems is motivated by the fact that there are hardly any systems older than 1 Gyr with solar metallicity.  Ideally, an age-metallicity relation should be employed for a smooth transition, which would be necessary to obtain an accurate SFH, covering the entire age range. For this work, where we restrict the SFH to the recent 100 Myr, the choice of only two metallicities and the chosen transition age does not affect the obtained results.

The age resolution of our SFH is primarily limited by the 0.1~dex age separation between successive isochrones. Given the sensitivity of the JWST CMDs to the entire age range used, stars at certain locations in the CMD could have an equal probability of belonging to completely different ages. One such location is to the left of the red giant branch (RGB), where the location in the CMD of stars of around a hundred million years (within the errors of interstellar extinction) coincides with the stellar population of several gigayears. The former corresponds to the core He-burning (CHeB) sequence of intermediate-age stars, whereas the latter corresponds to low-mass stars in their RGB phase. 

This degeneracy is naturally broken if we modify the statistical model described in \citet{superbubble} by adding a weight factor to account for logarithmic isochrone bins:

\begin{equation}
    W_i = \frac{\Delta t_i}{\sum_{i=0}^{N_{ISO}} \Delta t_i}.
\end{equation}    

Where $\Delta t_i$ is the age bin width of the i$^{\rm th}$ isochrone. $N_{\rm ISO}$ is the total number of isochrones. Due to the use of logarithmic age bins, $\Delta t_i$ becomes larger at older ages (e.g., 10~Gyr) and smaller at younger ages (e.g., 100~Myr). This, in turn, helps break the CHeB-RGB degeneracy.

Finally, after applying completeness cuts using m$_{50}$ values for each filter to the observed photometry, we run MCMC sampling on the statistical model to infer the $a_i$, which is converted to star formation rates (SFRs) using the procedure described in Section 4.3 of \citet{superbubble}. 

\begin{figure*}
    \centering
    \includegraphics[width=\linewidth]{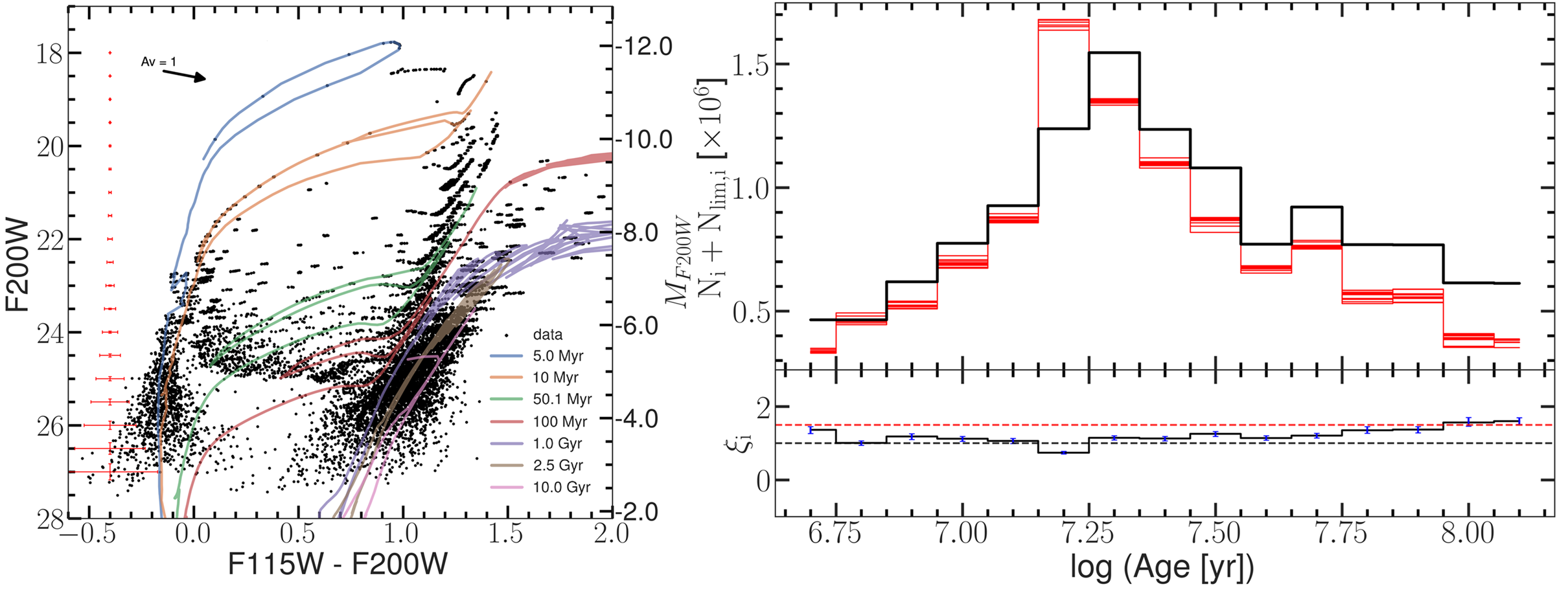}
    \caption{\textit{Left:} PARSEC simulated F115W$-$F200W vs F200W CMD with distance modulus and foreground extinction corresponding to NGC\,628 with a custom SFH. PARSEC+COLIBRI isochrones of selected ages, with the younger (log(Age yr$^{-1}$) $<$ 9) isochrones being metal-rich (Z = 0.02) than the older ones (Z = 0.002), are superposed. \textit{Right:} Bayesian SFH of the simulated CMD. The black line indicates SFH retrieved without completeness magnitude cuts. The red lines indicate retrieved SFH for CMDs with simulated completeness effects. The spread in the SFH originates from the Monte-Carlo sampling for completeness effects. The bottom panel shows the incompleteness correction, which is the ratio of the input number of stars and the retrieved number of stars. The dashed black and red horizontal lines indicate $\xi=1$ and $\xi=1.5$.}
    \label{fig:sim_CMD}
\end{figure*}

\subsection{Completeness, Incompleteness and Simulated CMDs}
\label{sec:incomp}
The ability to recover the SFH depends critically on the number of stars present in each age bin above the completeness limit. To assess the reliability of the recovered SFH, we test the recovery procedure on simulated CMDs. Synthetic CMDs are constructed by generating stellar populations of $10^{5}~M_{\sun}$ at each age bin within the range  6.7$\leq\log({\rm Age\,[yr]})\leq$10, with a step size of 0.1 dex, using the PARSEC web interface\footnote{\url{https://stev.oapd.inaf.it/cgi-bin/cmd_3.9}}. In this simple SFH simulation, all stars belonging to a logarithmic age bin have the same age corresponding to the midpoint of that bin. 

For building a specific SFH, the $10^{5}~M_{\sun}$ population is replicated as many times as required. For example, a $10^{6}~M_{\sun}$ population is obtained by adding the $10^{5}~M_{\sun}$ population ten times to the CMD. In this work, we adopt an SFH that guarantees at least 200 stars above the completeness limit in each age bin. This choice ensures adequate sampling across all ages while minimizing computational cost, in contrast to a constant star-formation rate case that would require unfeasibly large stellar populations at older ages due to their larger $\Delta t_i$.

Photometric errors are added based on the observed uncertainties, and completeness effects are incorporated using completeness curves for the three filters. Specifically, completeness is simulated by probabilistically removing stars, with survival probabilities assigned according to the Pritchet function~(Pr). The survival probability of each star is defined as:

\begin{equation}
    P_{{\rm Completeness}} = Pr(m_{\rm F115W})Pr(m_{\rm F150W})Pr(m_{\rm F200W})
\end{equation}

Using $P_{\rm Completeness}$, we generate several Monte Carlo (MC) samples of the synthetic CMDs for each bubble and grid cell. An example simulated F115W$-$F200W vs F200W CMD for one of the bubbles (Bubble \#0) is shown in Fig.~\ref{fig:sim_CMD}~(Left). In Fig.~\ref{fig:sim_CMD}~(Top Right), the comparison between the input SFH (black) and the recovered (red) SFH is shown for ages less than 100~Myr. The spread in the recovered SFH arises from the MC CMD samples and shows the variation in the offset. The ratio between the input SFH and recovered SFH is used as a correction factor (henceforth, incompleteness correction, $\xi$), which is calculated for each bubble and grid cell. The resulting incompleteness correction function for the illustrative region is shown in the bottom-right panel, where the horizontal black dashed line corresponds to a unitary correction factor. The correction factor ($\xi$) is less than 1.5 for ages less than 100~Myr for the illustrative region, as indicated by the horizontal red dashed line.

\begin{figure}
    \centering
    \includegraphics[width=\columnwidth]{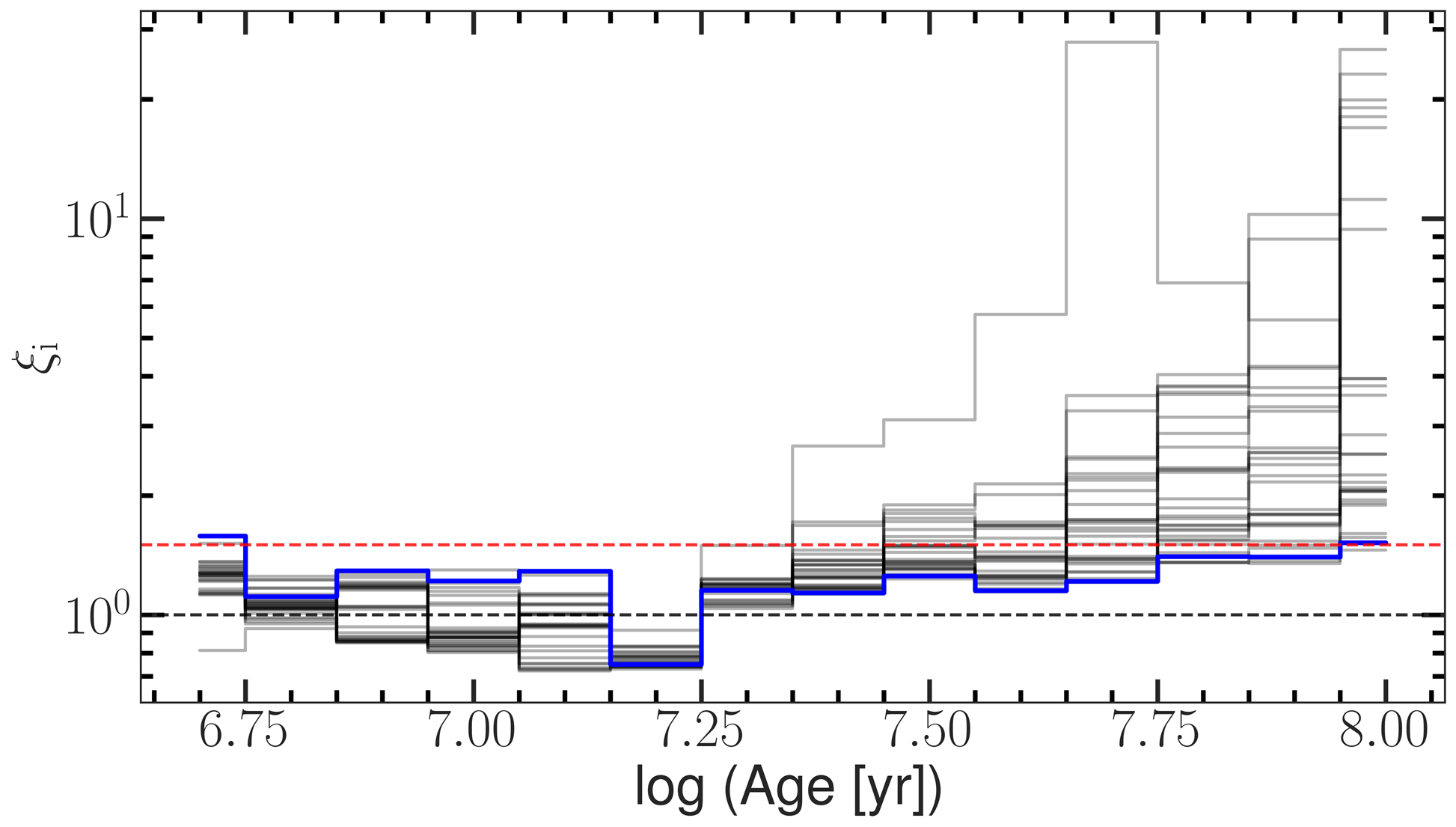}%
    \caption{Incompleteness correction factor for all 30 bubbles is indicated by grey step plots. The blue step plot highlights the incompleteness correction shown in Fig.\ref{fig:sim_CMD}~(Right). The dashed black and red horizontal lines indicate $\xi=1$ and $\xi=1.5$.}
    \label{fig:incomp_grid}
\end{figure}

The correction factor for all 30 bubbles is shown in Fig.~\ref{fig:incomp_grid}, which in general follows the tendency displayed for the illustrative region. However, for 9 bubbles $\xi$ rises above 1.5 for ages younger than 50~Myr. These are the bubbles that have completeness magnitude F200W$<$24.5 (see Fig.~\ref{fig:cmd} bottom), which results in the non-detection of the blue-loop stars of intermediate masses, which are the most abundant at ages beyond 50~Myr. The grid number (See Fig.~\ref{fig:grids} for the grid overlay on the F770W image) where the bubble is located, the youngest age at which $\xi>$1.5 for each bubble, and the completeness magnitude are given in columns 2, 3, and 4, respectively, in Table~\ref{tab:bubble_properties}.
For this study, we restrict our analysis to ages younger than 100~Myr, the age range where $\xi<$1.5, as we do not expect stellar populations older than this age inside the bubbles for the reasons described below.

In a systematic analysis of more than 1000 HI holes in 20 nearby galaxies, \citet{hi_hole2} found that the majority of the holes are consistent with being created by the feedback effects of massive stars. The smallest size of their holes is limited by the spatial resolution of 50~pc, which is similar to the minimum size of our big bubble sample. The dynamical ages of the majority of their holes are less than 100 Myr, with the few older holes lying beyond the optical radius of the galaxies. They also find the holes being more elliptical in regions of galaxies experiencing shear, illustrating that dynamical effects in the disks of spiral galaxies destroy the bubbles in timescales less than 100~Myr. The majority of the area covered by the JWST images in NGC\,628 experiences shear because they lie in the rising part of the rotation curve \citep{NGC628_sfr2}. Thus, we should be able to trace the populations responsible for the creation of MIRI-bubbles in NGC\,628 by constructing SFH up to 100~Myr.

In addition to the incompleteness correction, the simulated CMDs also revealed that a minimum of 10 massive stars (younger than 50~Myr) is required for a reliable recovery of the bubble SFH, which corresponds to at least 6500~M$_{\sun}$ of cluster mass for the Kroupa IMF and PARSEC isochrones. This is motivated by statistical considerations of 1 star per bin on average between 6.7$\leq \log(\rm{Age})\leq$ 7.7. While, in principle, even a single star would yield a formally defined SFH, such a solution would lack statistical robustness. In these low-number regimes, the Bayesian framework can produce an apparent SFH driven primarily by the underlying probability distribution functions, specifically by how individual stars map onto isochrones in the CMD, rather than by meaningful sampling of the underlying stellar population. By requiring a minimum of 10 young stars, we ensure that such stochastic effects do not dominate the inferred SFHs and instead reflect a minimally constrained, physically interpretable distribution of recent star formation.

\begin{figure*}
    \centering
    \includegraphics[width=\linewidth]{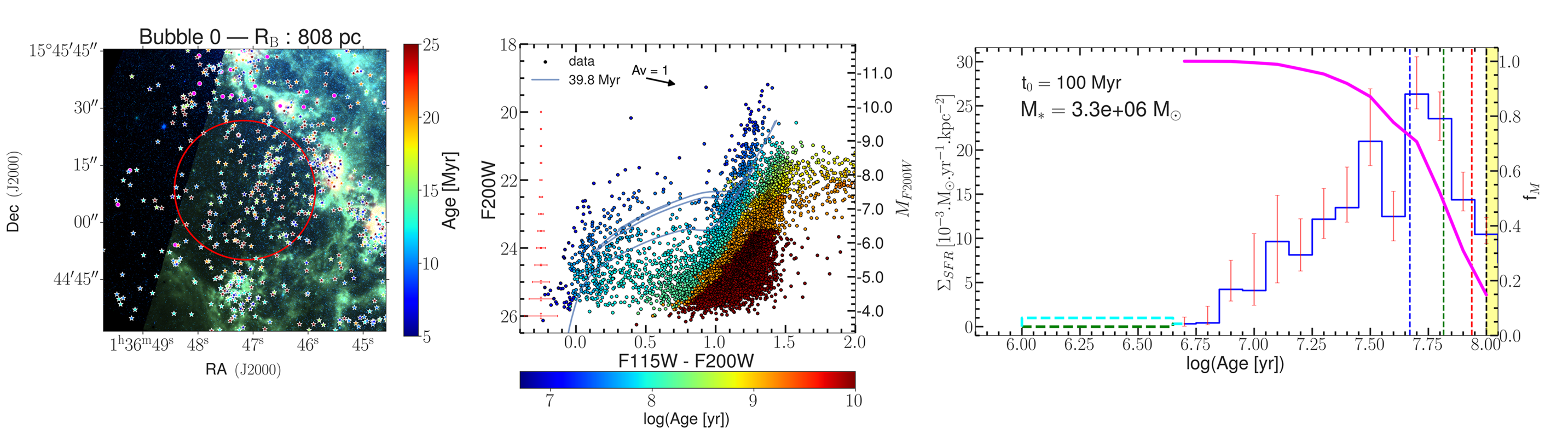}
    \includegraphics[width=\linewidth]{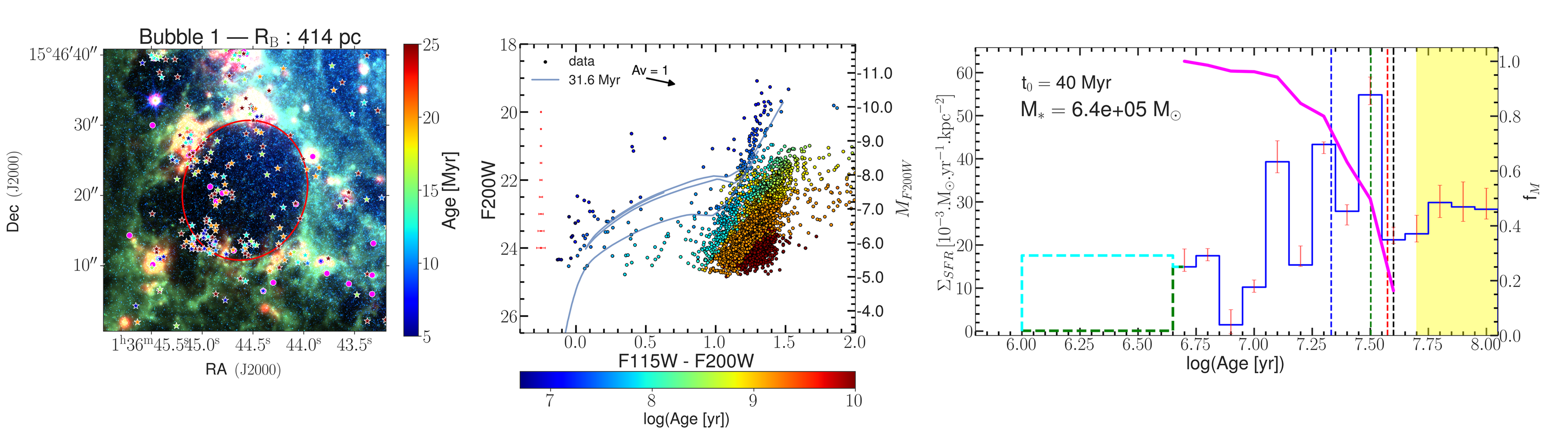}
    \includegraphics[width=\linewidth]{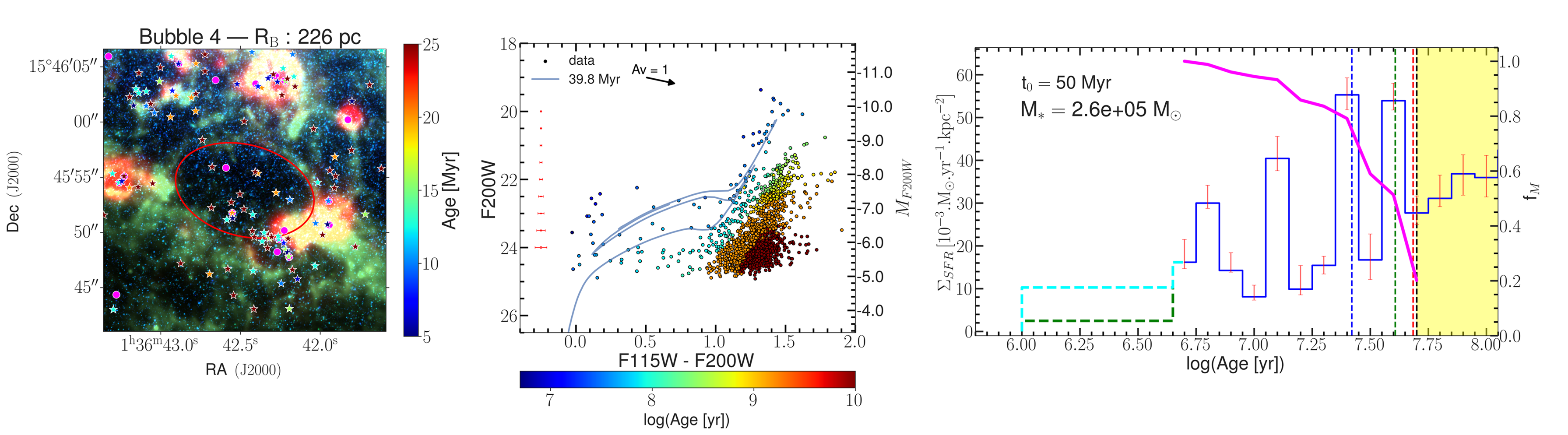}
    \includegraphics[width=\linewidth]{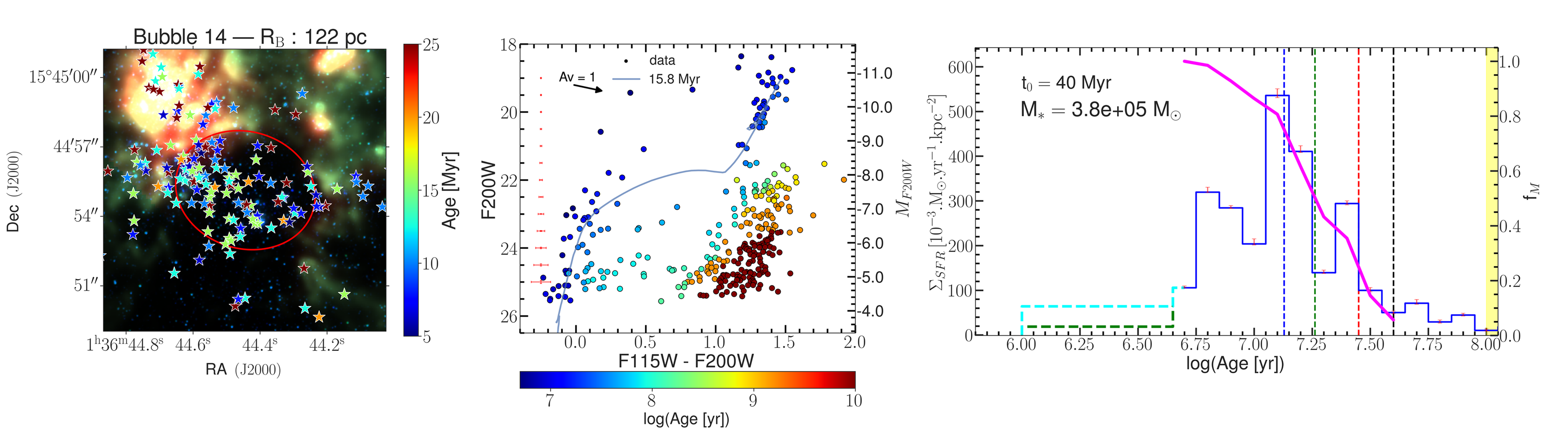}
    \caption{\textit{Left:} Color-composite image of a bubble using PHANGS-MUSE/H$\alpha$ (red), NIRCam/F200W (green), NIRCam/F115W (blue), and MIRI/F770W (light green). The red ellipse encloses the inscribed bubble. Filled star symbols denote stars younger than 25~Myr, with colors indicating ages derived from Bayesian isochrone fitting. Filled circles show PHANGS-HST clusters. \textit{Middle:} F115W$-$F200W vs.~F200W CMD, where the color represents the best-fitting stellar ages from Bayesian isochrone fitting. \textit{Right:} The blue step plot shows the star formation rate density, $\Sigma_{\rm SFR}$, of the excess stellar population inside the bubble. This excess is obtained by subtracting from the bubble $\Sigma_{\rm SFR}$ an average $\Sigma_{\rm SFR}$ of the disk population at the same galactocentric distance as the bubble. The yellow region indicates the age beyond which the derived $\Sigma_{\rm SFR}$ is uncertain due to a relatively large incompleteness correction ($\xi>1.5$). The black, red, green, and blue dashed lines mark $t_{0}$, $t_{25}$, $t_{50}$, and $t_{75}$, respectively. The magenta curve shows the normalized cumulative mass function integrated from $t_{0}$. The green and cyan dashed step plots indicate the SFR within the bubble and in its shell, respectively, estimated from the dust-corrected H$\alpha$ luminosity in PHANGS-MUSE maps.}
    \label{fig:bubble_SFH}
\end{figure*}

\subsection{Star formation history of the bubble population}
 
We show a color-composite image of all the big bubbles\footnote{Plots for the selected five bubbles are shown in the main text. Similar plots for other bubbles are provided in the supplementary material.} and the JWST CMD is shown in the left and middle panels of Fig.~\ref{fig:bubble_SFH}, respectively. The color-composite image contains four filters that enable tracing stellar continuum (F115W  and F200W as blue and green components, respectively), ionized gas (MUSE-H$\alpha$ in red), and PAH emission (F770W as light green). The contrast of the F770W image is adjusted so that the shell boundary of the selected bubble, which is depicted by a red ellipse, is noticeable. It may be recalled that a bubble is considered big if it contains at least 10 stars younger than 50~Myr. The stars in the left and middle panels are color-coded to differentiate their ages, as indicated by the color bar adjacent to each panel. The isochrone approximately representing stars younger than 100~Myr is overplotted in the middle panel, with its age indicated in one of the corners.  The isochrone is a poor fit to the data in all cases, illustrating the presence of multiple-age populations over the past 100~Myr inside the bubble. The SFH plot on the right panel contains many derived quantities, which are explained in the next subsection.

The primary goal of deriving the SFH of bubbles is to determine the age of the stellar population likely responsible for their formation. Some disk stars unrelated to the formation of the bubble are also expected to be seen projected onto the bubble area. In order to account for these stars, we characterize the SFH of the disk at the same galactocentric radius as that of the bubble. For this, we compute the SFH in each of the 90 grids of 24$\times$24 arcsec$^2$ area. For each bubble, we define a subjacent disk SFH as the median of the SFHs of all the azimuthal grids at the same galactocentric distance, omitting the grid (or grids if the bubble encompasses more than one grid) that contains the bubble. The grid-to-grid dispersion of the SFH is taken as the error on the subjacent disk SFH (See Appendix~\ref{sec:sub_disk} for more details on subjacent disk estimation). To account for the difference in areas of the bubble and the grids used for sampling the disk SFH, we convert the SFR to SFR density by dividing both the bubble and subjacent SFR by their respective areas in kpc$^2$ ($\Sigma_{\mathrm SFR}$). We subtract this baseline to obtain the excess populations inside the bubble, which must be responsible for the creation of the bubble.
Following the discussion in the previous subsection, we limit our SFH analysis to the last 100~Myr. 

In the right panel of Fig.~\ref{fig:bubble_SFH}, we show the $\Sigma_{\rm SFR}$ of the excess population of stars detected in the bubble (blue) over the last 100~Myr. The plotted SFH already takes into account the correction $\xi$ for both the bubble and disk stars. The error bars are obtained as the quadratic sum of Poissonian errors for the stars inside the bubble and azimuthal dispersion in the SFH of the disk at the same galactocentric distance as the bubble. 
In all cases, there is a positive excess after taking into account the error bars, especially at smaller ages. This excess confirms the presence of recently formed stars inside the bubble. We carefully analyzed this excess to find out the age of the population that most likely triggered the formation of the bubble. We define it as the oldest age $t_0$, above which the excess is zero after taking into account the errors on the determined SFH curves. In a few cases (9) that lie in the inner part of the galaxy, where the $\xi$ reaches values above 1.5 at age $<t_0$, we fix the $t_0$ as the age at which $\xi$=1.5, which is a lower limit to the age of the bubble population. This can be inferred from a comparison of columns 5 and 3 of Table~\ref{tab:bubble_properties}.

We did not find any bubble that required isochrones younger than $\sim$5~Myr. This is due to a combination of the following two effects:
1)~Massive main-sequence stars have large bolometric corrections in the JWST filters, which take them below our limiting magnitude. 2)~Most of the current star formation (SF) is in the shell, which is just outside our ellipse boundaries used for SFH analysis. We used the MUSE H$\alpha$ fluxes inside the ellipse and in the shell to obtain the respective $\Sigma_{\rm SFR}$(young). We display the resulting values as green and cyan-colored horizontal lines that fill the gap in the SFH obtained from the resolved populations. The SFH within the ellipse is nearly zero in all cases, indicating the absence of current SF within the bubble. On the other hand, the shell is active in the current SF in all cases. The SFR of the shell is almost identical to the average of the SFH in the first few bins obtained from the resolved stellar populations. This agreement of the SFR obtained from completely different approaches is heartening and validates the assumptions used in both these methods. The agreement also suggests that the shell SF is the continuation of the SF that was triggered inside the bubble.

\subsection{Derived quantities}

A quick glance at the $\Sigma_{\rm SFR}$ for each region reveals that the SFH is not peaked at a single bin, or equivalently, all the stars do not crowd around a single isochrone, as expected if the bubble was created by a single event of SF with no SF posterior to that event. Instead, the general trend is a more continuous SF of varying rates starting from $t_0$, the age of the precursor event that triggered the bubble formation, up to the last 5~Myr. As we illustrate later on in this section, stars from older events are distributed throughout the bubble, whereas stars from more recent events are closer to the shell. In order to obtain more insight into the SF inside the bubble, we obtain the growth rate of stellar mass inside the bubble. We start by constructing a cumulative mass fraction (f$_{\mathrm M}$) by integrating the SFH from t = t$_0$ to t = 5~Myr (from past to present) and normalizing it to the integrated total mass (M$_{*}$;  bubble mass). We define t$_{25}$, t$_{50}$, and t$_{75}$  as the ages when the bubble formed its first 25\%, 50\% and 75\% of this mass, respectively. The SFH of the bubble and the disk stars, t$_0$, t$_{25}$, t$_{50}$, and t$_{75}$, and the cumulative mass function for each bubble are all shown in the right panel of Fig.\ref{fig:bubble_SFH}.  These quantities are tabulated in columns 5 to 8 of Table~\ref{tab:bubble_properties}, and M$_{*}$ is tabulated in column 9.

\section{Discussion}

Having obtained useful physical quantities of the bubble population for a sample of 30 big bubbles, we now compare these properties with those expected theoretically from the massive-star-driven bubbles.

\subsection{Are the bigger bubbles older?}
\begin{figure}
    \centering
    \includegraphics[width=\columnwidth]{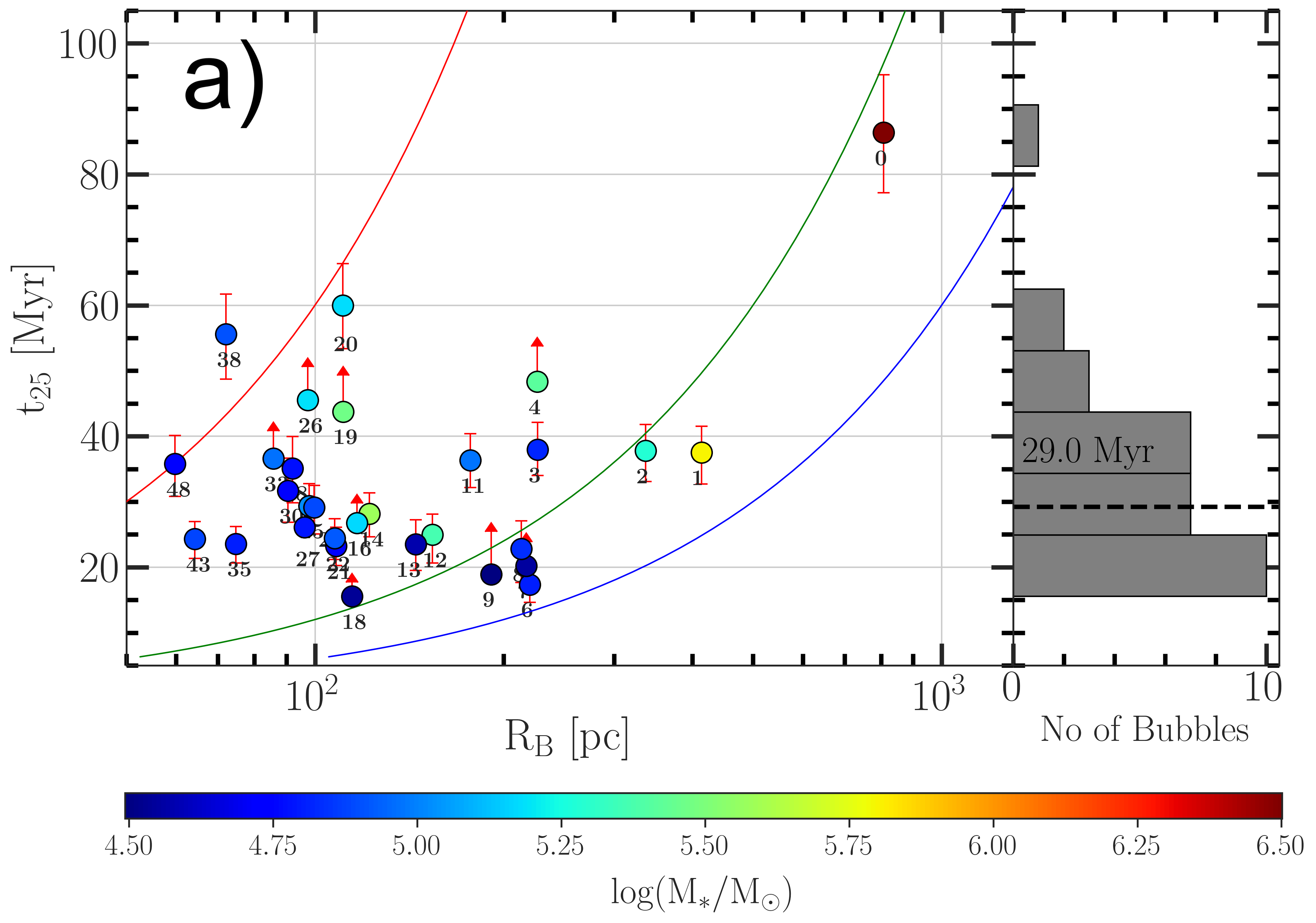}
    \includegraphics[width=\columnwidth]{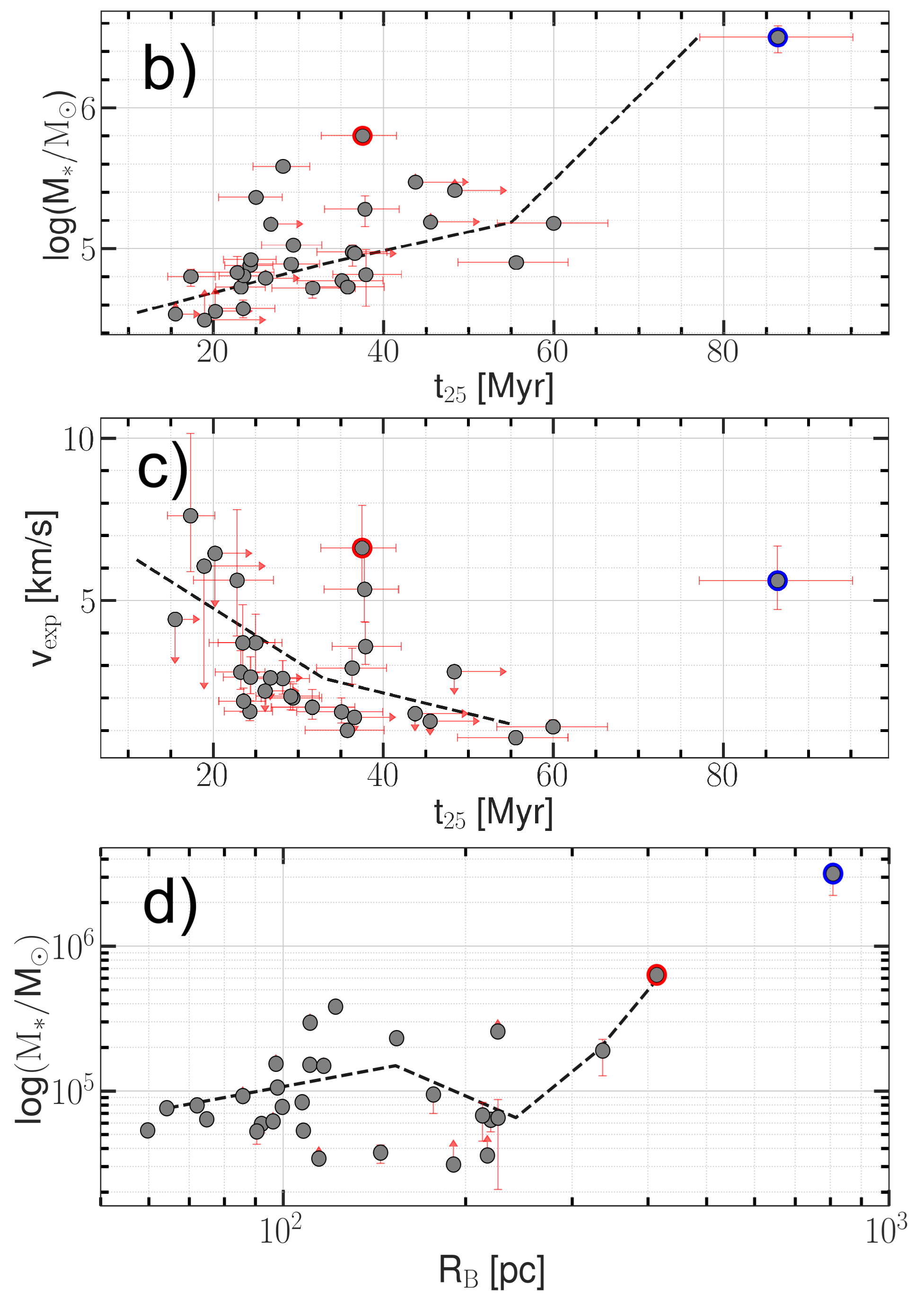}
    \caption{\textit{(a)} The left panel shows t$_{25}$ vs bubble radius. Filled circles indicate the primary bubbles. The red, green, and blue lines represent constant expansion speeds of 1~km/s, 5~km/s, and 10~km/s, respectively. The color axis indicates the mass accumulated since t=t$_0$. The right panel shows the distribution of t$_{25}$ with the black dashed line indicating the median age. \textit{(b)} Stellar mass of stars formed inside bubble (M$_{*}$) vs t$_{25}$. \textit{(c)} Expansion velocity as a function t$_{25}$. \textit{(d)} Stellar mass of stars formed inside the bubble vs the circularised radius of the bubble. The red and blue outlined circles indicate the Phantom void and HI-hole in the plots.}
    \label{fig:t25}
\end{figure}

\begin{figure*}
    \centering
\includegraphics[width=1.5\columnwidth]{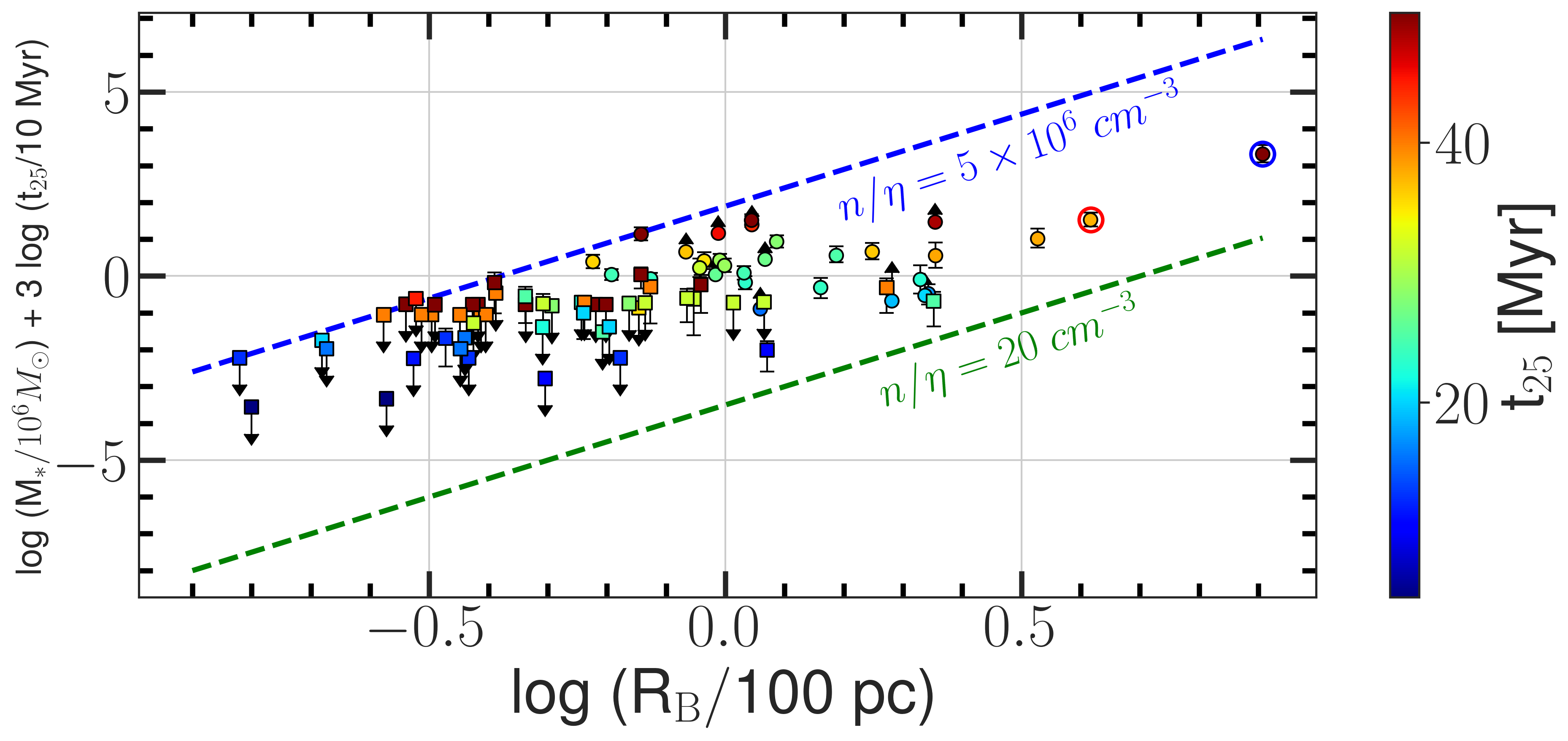}
    \caption{Fundamental plane of the stellar feedback-driven bubbles. The plot includes not only the big bubbles (filled circles), but also smaller bubbles (inverted triangles) that contain at least one star younger than 50~Myr.
    The dashed lines that enclose all the observed points represent Eq.~\ref{eq:energy} corresponding to the annotated values of \textbf{$n/\eta$}. The observed points are color-coded based on their ages, following the color bar to the right. The red and blue encircled points indicate the Phantom Void and HI-hole, respectively. See text for details.}
    \label{fig:energy}
\end{figure*}

In the scenario of bubbles formed from stellar feedback from multiple generations of star formation, continuously injecting momentum and energy into the ISM, we should expect larger bubbles to have older stellar populations. Additionally, for a bubble expanding in a medium of constant density, the bubble radius is expected to grow+ following the equation: 
$R_{\mathrm{B}} = \frac{5}{3} v_{\mathrm{exp}}\ t$ 
\citep{v_exp1,v_exp2}. 

In Fig.~\ref{fig:t25}~(a), we show $t_{25}$, taken as a proxy for the bubble age, against bubble radius. Most of the bubbles have ages between 20 and 60~Myr with a median of $t_{25}$ of $\sim$30~Myr. There is no relation between the two plotted quantities, implying that, in general, bigger bubbles are not systematically older. However, if we restrict the analysis to bubbles of $R_{\rm B}<$100~pc, we see a tendency for the oldest of the bubbles to be larger. On the other hand, there are some 100~pc-sized young bubbles. For bubbles with $R_{\rm B}>$100~pc, the upper age boundary decreases with bubble size. This could be because the bubble size depends on other quantities, such as the non-constant expansion velocity and the amount of injected kinetic energy, which is proportional to the stellar mass formed since $t_0$. To illustrate the multi-parameter dependence of the bubble radius, we overlay the plot with lines of constant expansion velocity and color-code each point based on the accumulated stellar mass. It can be inferred from the plot that a range in expansion velocities between 1 and 5~km~s$^{-1}$ can produce the observed scatter at any given age. Thus, expansion velocity, not age, is the fundamental controller of the size. 

There is a tendency for older bubbles to harbour more massive stellar populations at fixed expansion velocities. To investigate this tendency, in panel (b) of Fig.~\ref{fig:t25}, we plot the stellar mass formed within each bubble as a function of the age of its oldest stellar population. 

A general trend is evident in which older bubbles tend to host more massive stellar populations. The stellar mass is not formed with a single burst, and instead accumulates over time with stars forming continuously. 

The size being controlled by expansion velocity rather than stellar mass or age is intriguing. In order to understand this nature, we explore the expansion velocity evolution by calculating $v_{\rm exp}$ for each bubble, assuming linear growth. The resulting  $v_{\rm exp}$ is plotted as a function of age in panel (c) of Fig.~\ref{fig:t25}. A tendency for the older bubbles to have lower $v_{\rm exp}$ can be inferred from this figure, suggesting decelerating bubbles. 

In Fig.~\ref{fig:t25}~(d), the mass of stars projected inside the bubble is plotted against the circularized radius of the bubble, and we find that bigger bubbles are more massive, albeit with a substantial spread in the relation. The scatter is possibly due to a combination of reasons, including selection effects, feedback efficiency, and the local gas density of each bubble. However, the increase in bubble size with an increase in stellar mass of stars formed projected inside the bubble is consistent with bubbles being created due to accumulated stellar feedback.

From the above analysis, we draw the following inferences. Bubble size grows linearly with time until it reaches a size of $\sim$100~pc, which is the typical scale height of the interstellar gas in the disk. This corresponds to the spherical bubbles in the energy-conserving phase. When the bubble size exceeds the scale height, it breaks open, blowing out the energy injected by the exploding supernovae. Such a bubble takes a cylindrical form, growing in diameter due to the momentum conservation, which makes the $v_{\rm exp}$ decrease with time as we observe. In this phase, energy injected from exploding supernovae is lost into the halo, and the size is controlled only by the momentum gained when the blowout happens.

\subsection{On the energetics of wind and supernova-driven bubbles}

From a theoretical standpoint \citep[][and references therein]{v_exp1,v_exp2}, the size of a bubble depends on the total feedback energy injected and the timescale over which this energy is dissipated. The mechanical energy driving the expansion is directly proportional to the total mechanical energy supplied by the underlying stellar population. The mechanical power liberated by the stellar population is given by

\begin{equation}
\begin{split}
L_{\mathrm{Mech}}^* 
&= \sum\limits_{i}^{n_{\mathrm{pop}}}
   \left(\dfrac{L_{\mathrm{kin}}}{M}\right)_{\rm SSP}\, M_*(t_i)
   = \left(\dfrac{L_{\mathrm{kin}}}{M}\right)_{\rm SSP} M_T \\
&= 5\times10^{40}
   \left(\dfrac{M_T}{10^6}\right)
   \,\mathrm{erg~s^{-1}},
\end{split}
\label{eq:L_star}
\end{equation}
where $L_{\mathrm{kin}}$ is the kinetic power of a stellar population of mass $M$ as calculated from single stellar population (SSP) models, $M_*(t_i)$ is the mass of the $i^{\rm th}$ population, and $M_T$ is the total stellar mass contributing to feedback.

The mechanical power contained in a spherical shell of radius $R_{\mathrm{b}}$ and swept-up mass $M_{\mathrm{sh}}$ can be written as
\begin{equation}
    L_{\mathrm{Mech}}^{\mathrm{gas}} =
    \left[\dfrac{7(9\gamma -4)}{27(\gamma -1)}\right]
    M_{\mathrm{sh}}
    \dfrac{v_{\mathrm{exp}}^3}{R_{\mathrm{b}}},
    \label{eq:L_gas}
\end{equation}
where $M_{\mathrm{sh}} = \dfrac{4\pi}{3} R_{\mathrm{b}}^3 nm_{\mathrm{p}} \mu$ for a spherical bubble, and $v_{\mathrm{exp}} = \dfrac{3}{5}\dfrac{R_{\mathrm{b}}}{t}$ \citep{stellar_feedback} for free expansion, and $\gamma$= 5/3 is the adiabatic index.
Equating equation~\ref{eq:L_star} and equation~\ref{eq:L_gas} yields
\begin{equation}
    L_{\mathrm{Mech}}^{\mathrm{gas}} = \eta L_{\mathrm{Mech}}^*,
\end{equation}
where $\eta$ is the efficiency factor that quantifies the fraction of mechanical energy from stellar feedback that is transferred to the gas, driving the bubble’s expansion. This relation can be rewritten as

\begin{equation}
\begin{split}
\log\left(\dfrac{M_T}{10^6 M_{\sun}}\right)
+ 3\log\left(\dfrac{t}{10~\mathrm{Myr}}\right)= {} \\
5\log\left(\dfrac{R_{\mathrm{b}}}{100~\mathrm{pc}}\right) 
+ \left[ \log\left(\dfrac{n}{\mathrm{cm^{-3}}}\,\dfrac{0.05}{\eta}\right) - 3.5\right],
\end{split}
\label{eq:energy}
\end{equation}
where $n$ is the interstellar medium (ISM) particle density. We have here used the $\left(\dfrac{L_{\mathrm{kin}}}{10^6M_{\sun}}\right)_{\rm SSP}=5\times10^{40}$ from STARBURST99 \citep{starburst99} for a solar metallicity single stellar population (SSP). Eq.~\ref{eq:energy} represents the fundamental plane for stellar feedback-driven bubbles.

In Fig.~\ref{fig:energy}, we plot the observed values for our sample of bubbles in the fundamental plane equation (Eq.~\ref{eq:energy}). The plot includes not only the big bubbles (filled circles) but also smaller bubbles (filled squares) that contain at least one star younger than 50~Myr. For the big bubbles, we use $M_*$ and $t_{25}$ from Table~\ref{tab:bubble_properties}, which are obtained from a detailed SFH analysis. For smaller bubbles, we use the median age of stars formed within the last 50~Myr inside the bubble as $t_{25}$ and the cluster mass required to produce the observed number of stars as M$_*$. These latter masses vary between 1000 and 15000~M$_{\sun}$. Due to stochastic IMF sampling and low-number statistics, the derived stellar masses for the smaller bubbles are potentially upper limits, because of which we indicate these masses as upper limits in the plot.

The dashed lines (slope=5) that enclose all the observed points represent Eq.~\ref{eq:energy} corresponding to the annotated values of $n/\eta$. 
The upper and lower boundaries of points correspond to dense regions with low mechanical feedback efficiency ($n/\eta=5\times10^6\,{\rm cm}^{-3}$) and the low-density regions with higher mechanical feedback efficiency ($n/\eta=20\,{\rm cm}^{-3}$). These parameter ranges encompass nearly all bubbles in our sample. This comparison illustrates that the models of stellar feedback triggering an expanding bubble in a uniform medium are able to reproduce the general demography of the bubbles seen on the MIRI images. In reality, the ISM in star-forming regions is far from a uniform medium, and the bubbles are not energy-driven once their size exceeds the disk scale height, which would make the properties deviate from the simple theoretical model presented above. Specifically, the bubble geometry changes from spherical to cylindrical shape, as can be seen in the detailed modelling of the Phantom Void by \citet{phantom_void_sim}. In fact, we see evidence for a break in the fundamental plane relation at $R_{\rm B}$=100~pc towards flatter slopes for larger bubbles, as expected for expanding bubbles that change from a spherical to cylindrical form. Hence, in summary, the bubble properties follow the fundamental plane relation expected for stellar feedback-driven bubbles.

\subsection{On the scenario of star formation in an expanding bubble}
\begin{figure}
    \centering
    \includegraphics[width=\columnwidth]{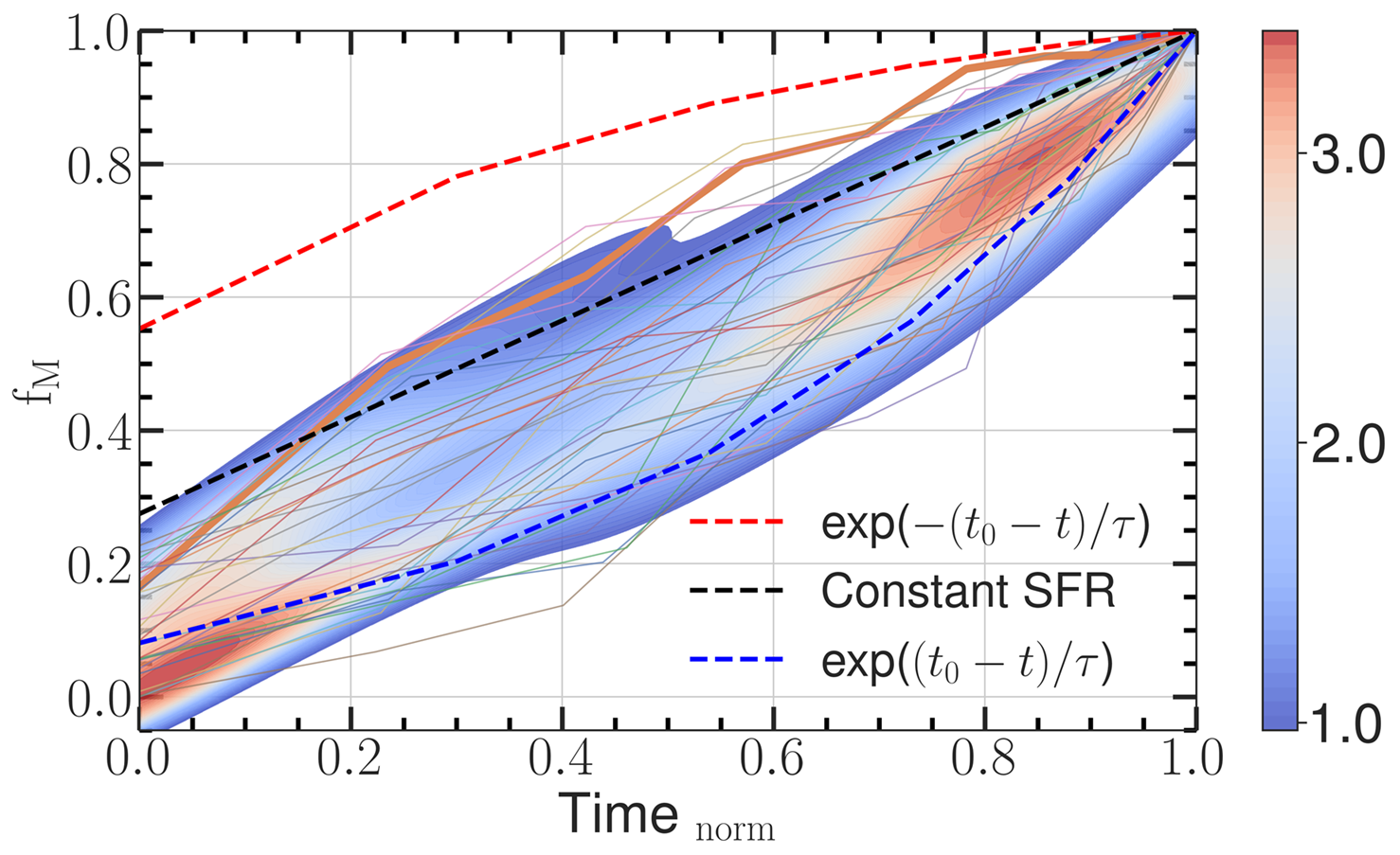}
    \caption{ Normalized cumulative mass fractions of bubbles. The red dashed curve indicates the case for exponentially decaying SFR with $\tau=10$~Myr, the black dashed curve indicates the case of constant SFR, and the blue dashed curve indicates the case for exponentially growing SFR with $\tau=10$~Myr. The thick orange line indicates the SFR of Phantom Void. The color map indicates the kernel density of f$_{\mathrm M}$s of the bubbles. See text for details.}
    \label{fig:cmfs}
\end{figure}

\begin{figure}
    \centering
    \includegraphics[width=\columnwidth]{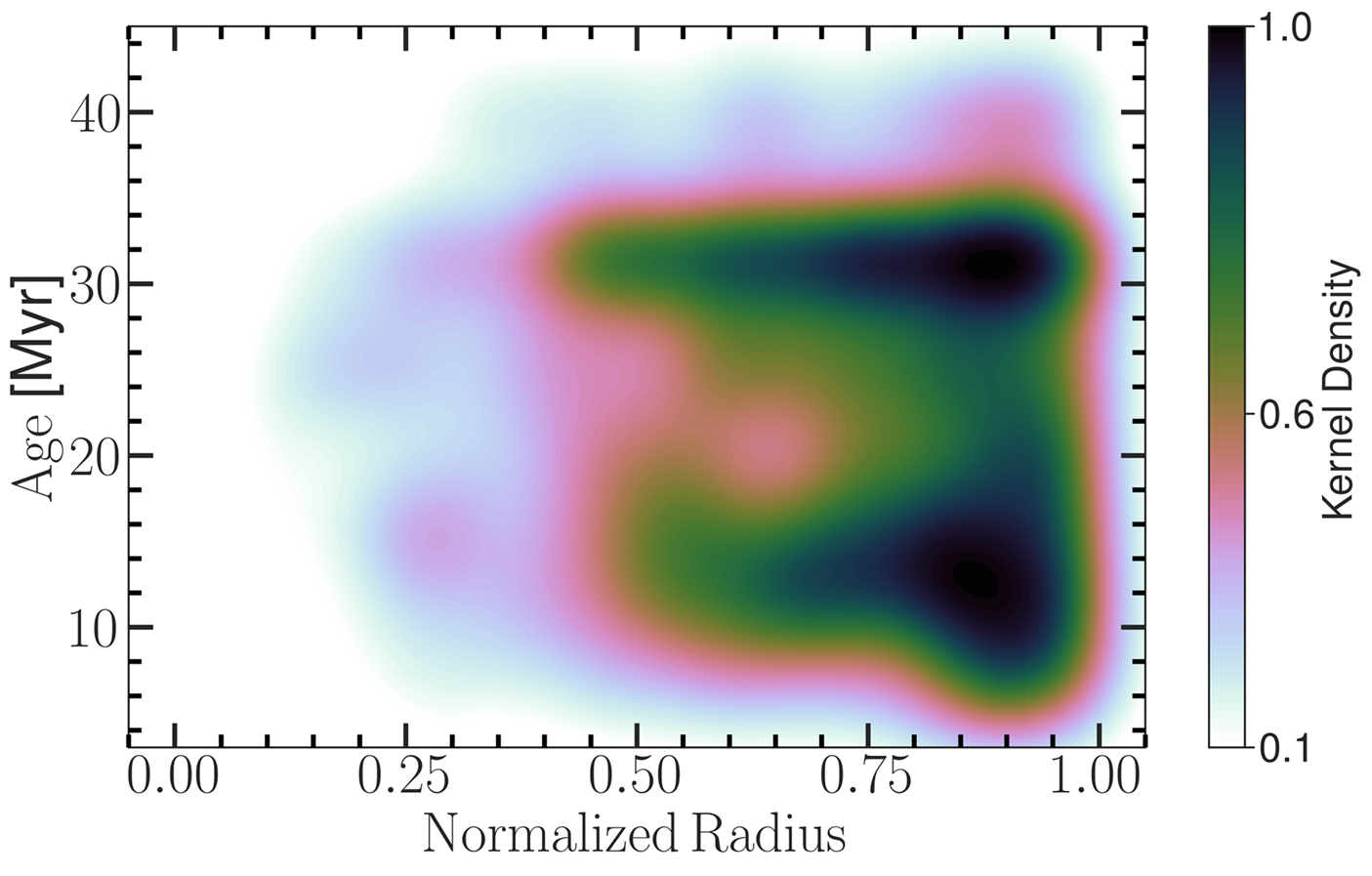}
    \caption{Best-fitting stellar ages as a function of distance of star with respect to the bubble centre, normalized with bubble radius, shown using the kernel density of the stars, for all bubble stars in our sample with age$<=$t$_0$ (Median age for small bubbles). }
    \label{fig:age_radius}
\end{figure}

Analysis of the SFH of our big bubble sample in the previous section showed that all bubbles enclose populations formed over an extended period of time, similar to what was found for H\,\textsc{i} holes in dwarf galaxies by \citet{bubble_hst_cmd} and \citet{bubble_hst_cmd2}, and more recently for the Phantom Void by \citep{superbubble}, all using HST CMDs. In the case of the Phantom Void, most of the stellar mass was formed at early times. In Fig.~\ref{fig:cmfs}, we show how the stellar mass was built up in each of the bubbles by plotting the fraction of accumulated stellar mass  f$_{\mathrm{M}}$ as a function of normalized time, where 0 corresponds to the epoch when the event of SF that triggered the formation of the bubble happened, and 1 corresponds to the present epoch. 

The observed f$_{\mathrm{M}}$ values (thin lines) are compared with those from models corresponding to a constant SFR (black), an exponentially declining SFR (red), and an exponentially rising SFR (blue). We construct the kernel densities of f$_{\mathrm{M}}$ at each normalized time to indicate the general behaviour of f$_{\mathrm{M}}$ as a function of time. The peak of the kernel density lies between the blue and black lines, implying a steady build-up of stellar mass after the initial event that triggered the bubble formation. 

The behaviour of the SFH of the majority of the bubbles suggests that a steadily increasing SFH is more successful in generating bubbles than events wherein most of the mass is formed in a single burst. This is consistent with the results of the simulations of \citet{superbubble_mult_gen}, who found a greater chance of survival of bubbles if the energy deposition is steady over time instead of an instantaneous burst. On the other hand, the curve for the Phantom Void (thick orange line) lies above the peak of the kernel density, following the form of the exponentially decreasing SFR. It is likely that the first generation of stars did not lead to the creation of the Phantom Void. Instead, the current size of the Phantom Void is maintained by the subsequent generations of stars. Its location in the inter-arm region allowed its growth without encountering any neighbouring expanding bubbles.

\citet{superbubble} proposed a scenario of star formation in which successive generations of stars are formed in the shell of the expanding bubble. The characteristic signature of such a scenario is the spatial segregation of stellar ages, with the youngest populations residing at the inner boundary of the current shell position, $R_{\rm B}$. Having obtained the ages of all stars inside the bubble, we now investigate whether such a scenario holds for our bubble sample. In Fig.~\ref{fig:age_radius}, we present the distribution of stellar ages within the bubbles as a function of elliptical distance\footnote{Elliptical distance is the effective radius of the ellipse that passes through the star with the same parameters as the ellipse inscribing the bubble.} from the bubble centre, with the distance for each bubble normalized by the bubble's effective radius. The values are plotted as the kernel densities, with darkest regions being the densest. The youngest stars ($<$15~Myr) are systematically located close to the shell (75–95\% of the bubble radius), with hardly any star in the inner 40\% of the radius. On the other hand, relatively older stars (25--45~Myr) are distributed beyond 25\% of the normalized radius. The trend of recent generations of stars forming at outer radii continues to the present day, with the shell being the currently active site of star formation. This can be inferred from the H$_\alpha$-derived SFR (cyan dashed line) of the shell shown in the right panels of Fig.~\ref{fig:bubble_SFH}. As pointed out previously, the shell $\Sigma_{\mathrm{SFR}}$ derived from H$_\alpha$ over the last 5~Myr is comparable to the average over the last few age bins derived from the resolved stellar populations. 

The presence of active star formation in the shell and the spatial segregation of stellar ages inside it support the star-formation in expanding bubble scenario proposed to explain the trend seen in the Phantom Void. It is interesting to note that the spatial segregation of ages is present independent of whether the SFR in the expanding shell increases or decreases with time. Thus, spatially segregated star formation seems to be a natural consequence of shell expansion.

\subsection{On the possibility of compact clusters as precursors of bubbles}
\begin{figure}
    \centering
    \includegraphics[width=\columnwidth]{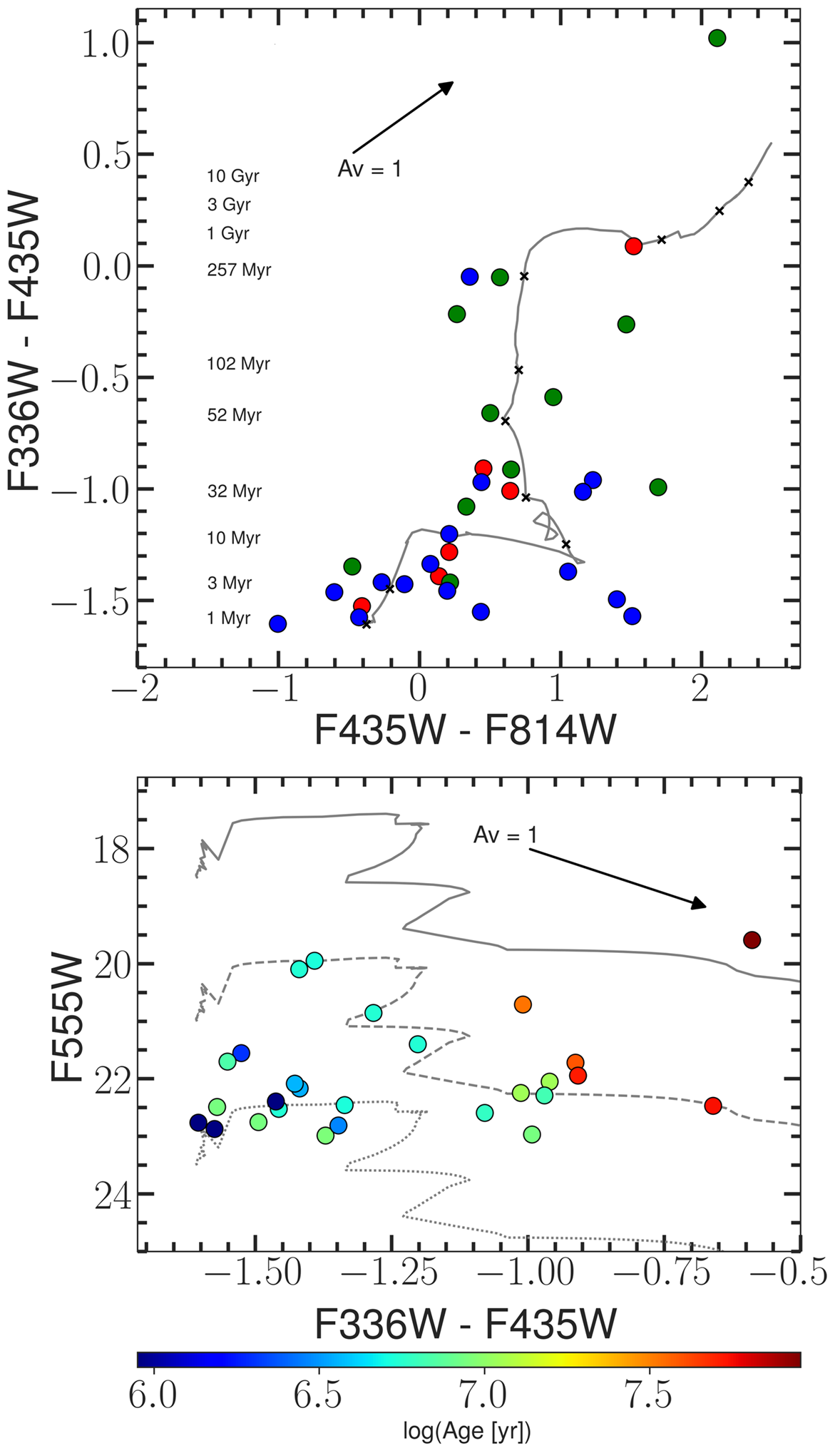}
    \caption{\textit{Top:} HST Color-color diagram of cluster candidates inside bubbles. Red, green, and blue filled circles indicate PHANGS cluster classes 1, 2, and 3, respectively.  Classes 1 and 2 are compact clusters, and Class 3 is for compact associations based on PHANGS-HST human cluster classifications. The grey curve indicates the SSP models from BC03, with black crosses indicating selected ages. The black arrow indicates the reddening vector for A$_{\mathrm{V}}$=1. \textit{Bottom:} HST CMD of cluster candidates inside bubbles. The clusters are color-coded by approximate ages obtained from the color-color diagram. The dotted, dashed and solid curves indicate BC03 SSPs of mass 10$^3$, 10$^4$, and 10$^5$~M$_{\sun}$, respectively.}
    \label{fig:clusters}
    
\end{figure}

\begin{figure}
    \centering
     \includegraphics[width=\columnwidth]{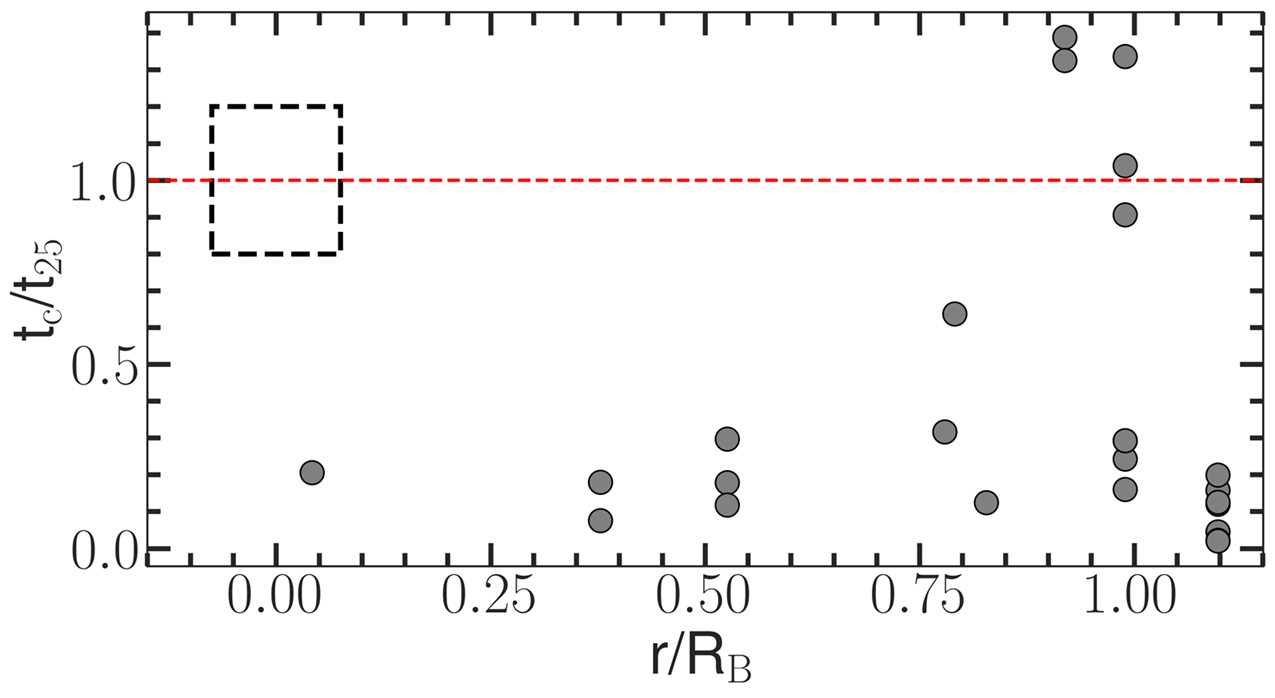}
    \caption{ The ratio of the cluster age ($t_c$) to the bubble age ($t_{25}$) as a function of the cluster distance from the bubble center, normalized by the bubble effective radius. The red dashed line indicates the locus where $t_c = t_{25}$. The black dashed rectangle highlights the region in which potential precursor clusters are expected to reside.}
    \label{fig:tc}
\end{figure}

\begin{figure*}
    \centering
    \includegraphics[width=\linewidth]{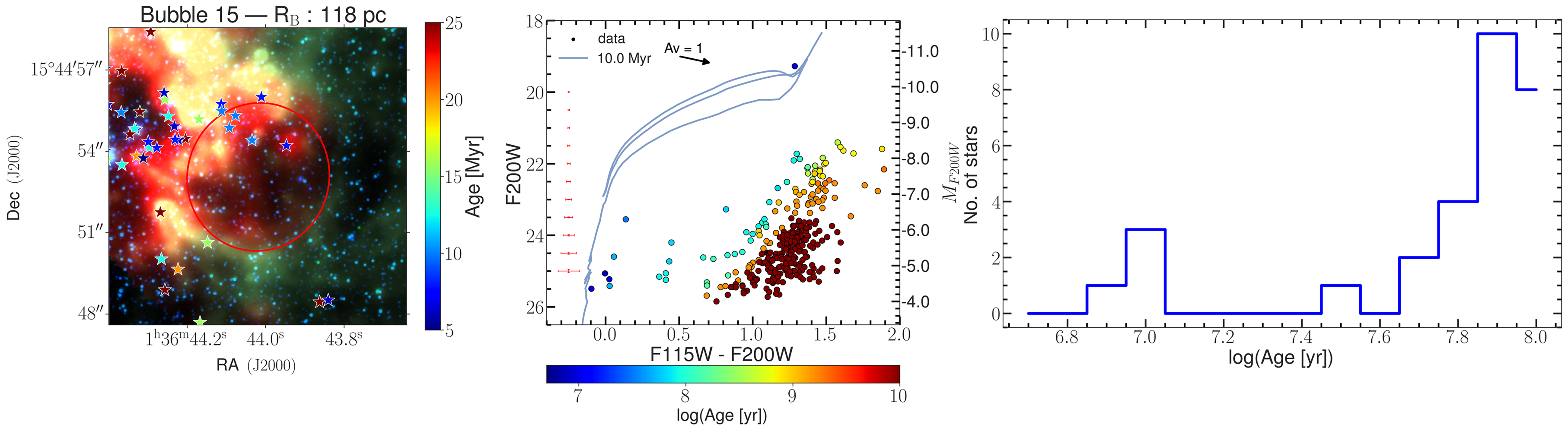}
    \includegraphics[width=\linewidth]{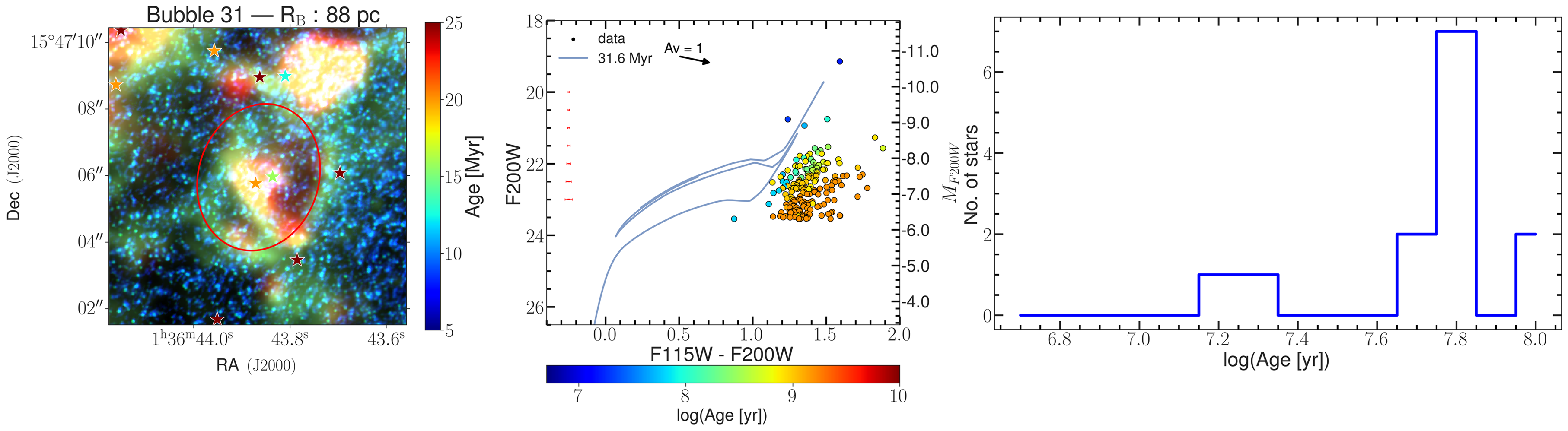}
    \includegraphics[width=\linewidth]{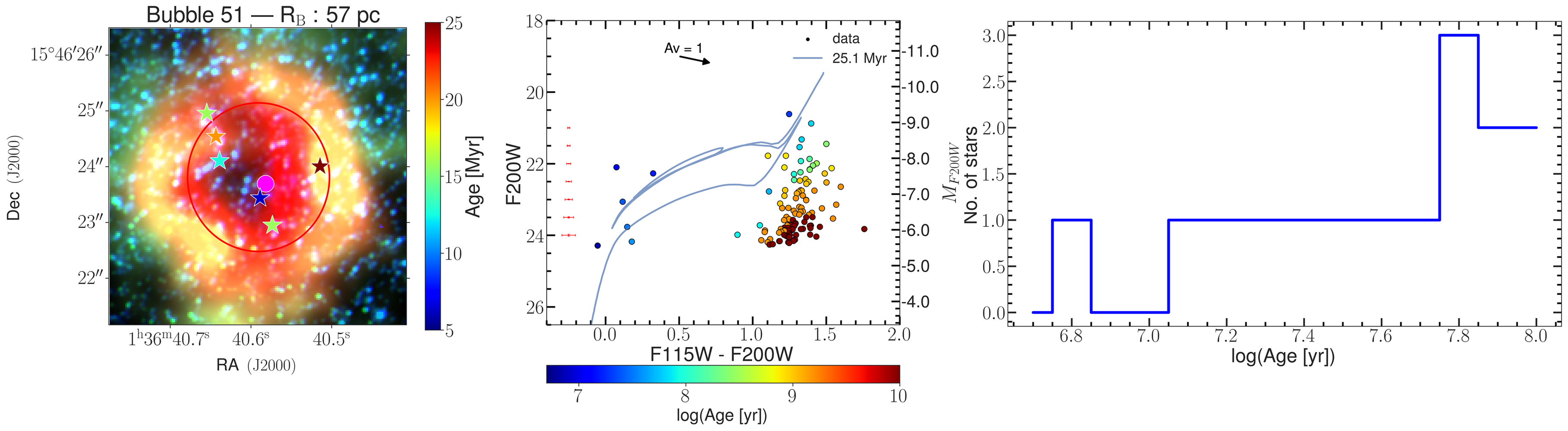}
    \includegraphics[width=\linewidth]{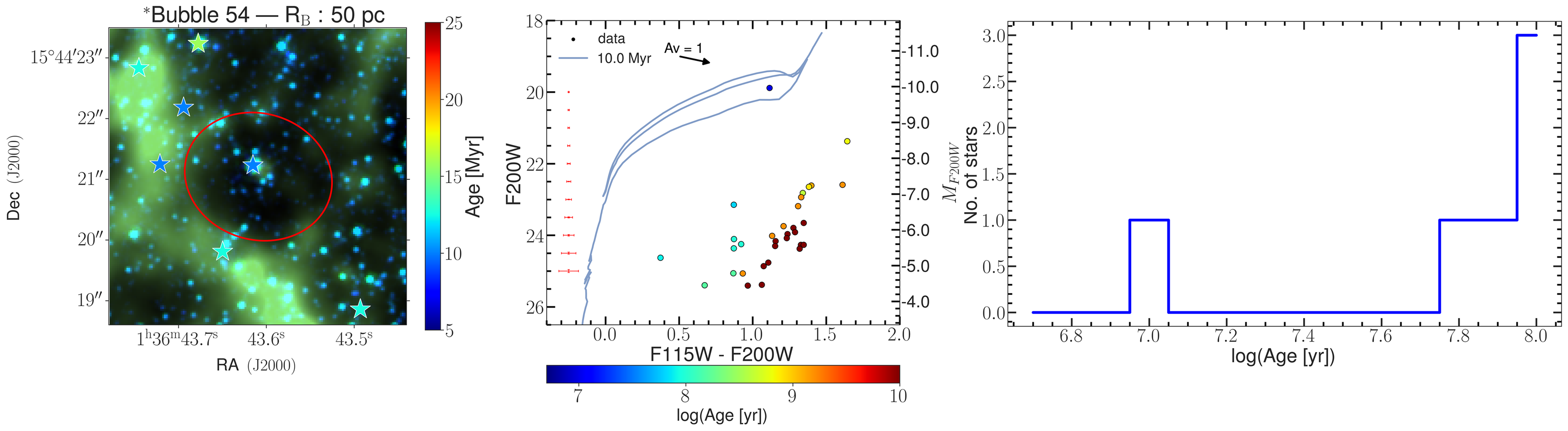}
    \caption{Bubbles with at least 1 star and at most 10 stars, younger than 50~Myr. \textit{Left:} Color-composite image of a bubble using PHANGS-MUSE/H$\alpha$ (red), NIRCam/F200W (green), NIRCam/F115W (blue), and MIRI/F770W (light green). The red ellipse encloses the inscribed bubble. Filled star symbols denote stars younger than 25~Myr, with colors indicating ages derived from Bayesian isochrone fitting. Filled magenta circles show PHANGS-HST clusters. \textit{Middle:} F115W$-$F200W vs.~F200W CMD, where the color represents the best-fitting stellar ages from Bayesian isochrone fitting. \textit{Right:} No of stars vs log (Age) of stars within the bubble. $^*$ Indicate bubbles for which PHANGS-MUSE data were not available.}
    \label{fig:bubble_SFH_small}
\end{figure*}

\begin{figure*}
    \includegraphics[width=0.33\linewidth]{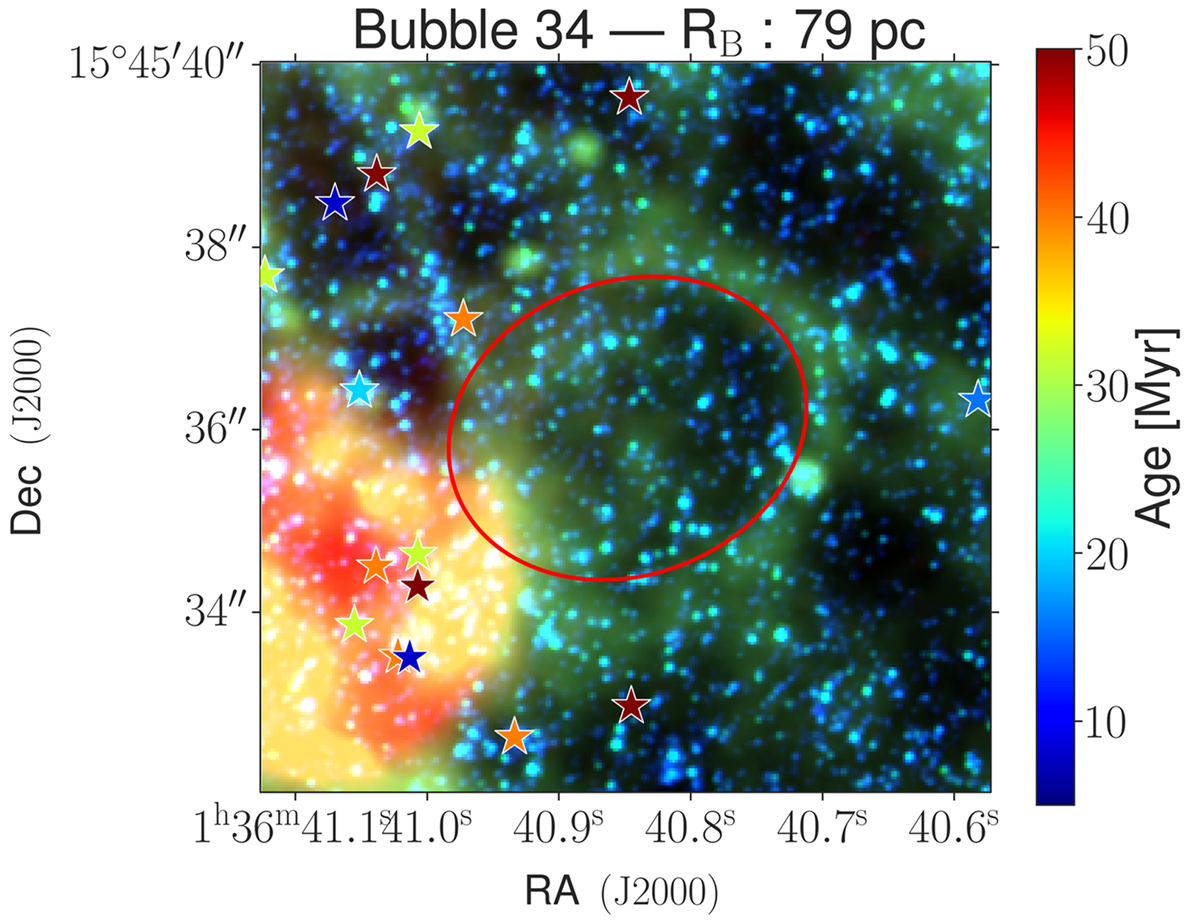}%
    \includegraphics[width=0.33\linewidth]{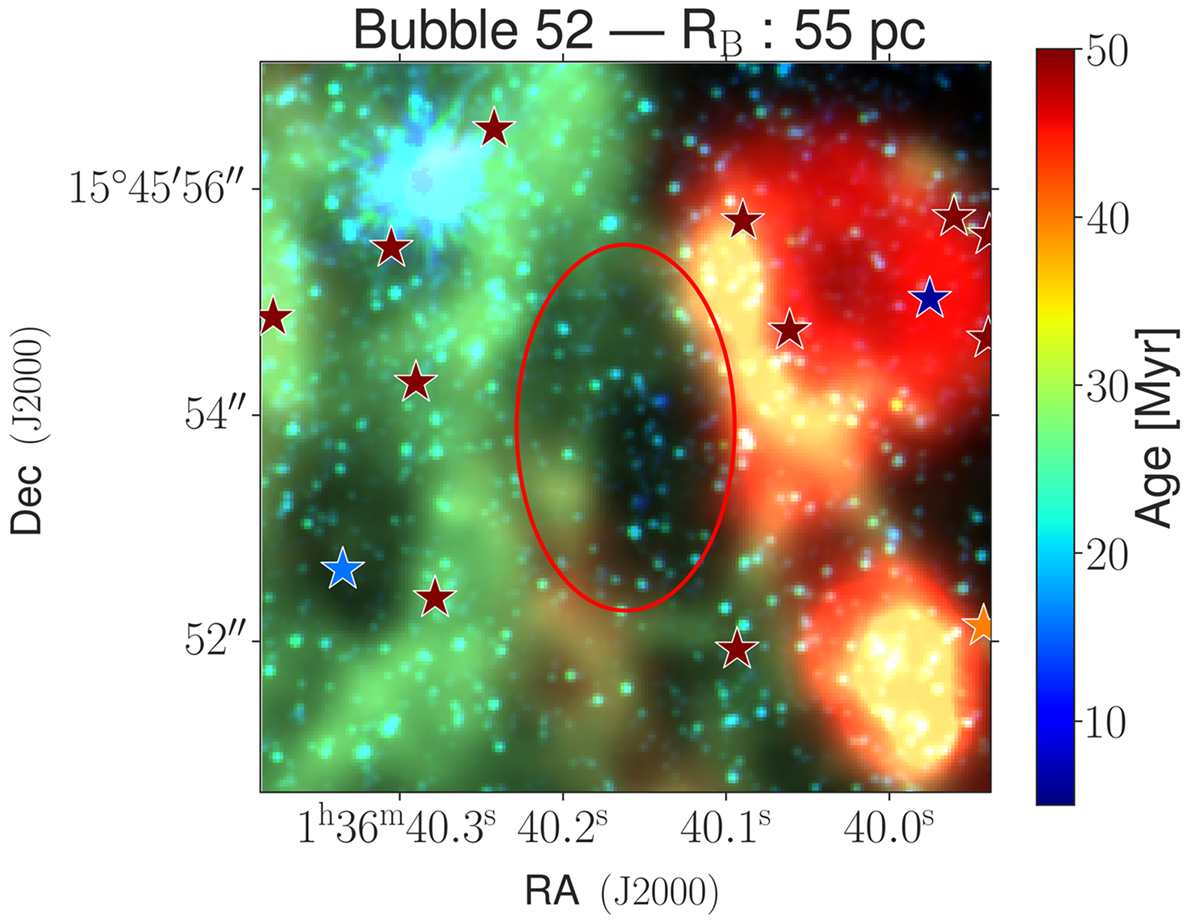}%
    \includegraphics[width=0.33\linewidth]{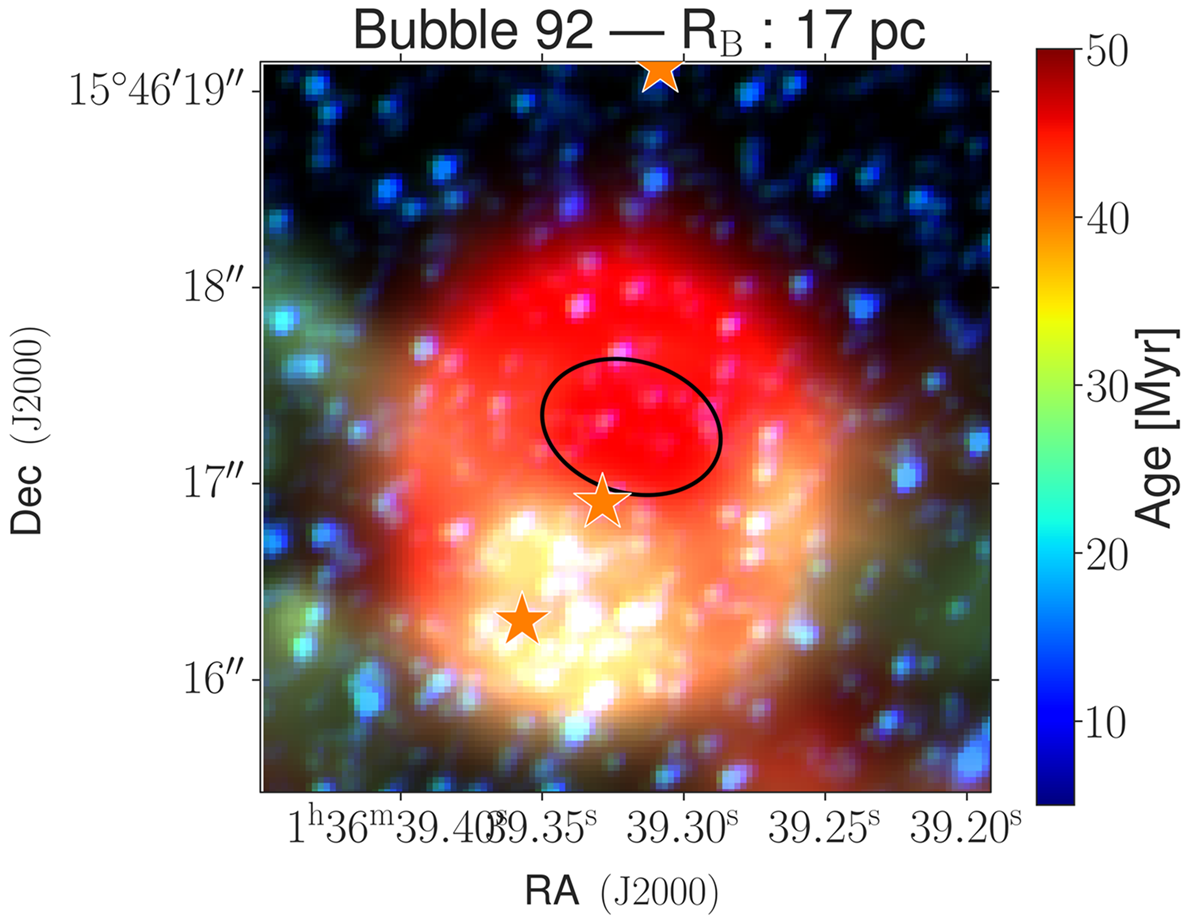}
    \caption{Bubbles with no stars younger than 50~Myr. Color-composite image of a bubble using PHANGS-MUSE/H$\alpha$ (red), NIRCam/F200W (green), NIRCam/F115W (blue), and MIRI/F770W (light green).  The red ellipse (black circle in the right panel) encloses the inscribed bubble. Filled star symbols denote stars younger than 50~Myr, with colors indicating ages derived from Bayesian isochrone fitting.}
    \label{fig:bubble_SFH_vsmall}
\end{figure*}

The resolved stellar population analysis we have carried out assumes that the high-mass end of the precursor stellar population is resolved into stars. This was motivated by the visual appearance of massive SG stars inside all big bubbles, illustrated in Fig.~\ref{fig:bubble_sample}. Only unresolved sources whose intensity profiles are consistent with the PSF profile are retained in the catalogue, which does not exclude the possibility of the presence of marginally resolved compact clusters inside the bubble. We utilize the PHANGS–HST cluster catalogue \citep{phangs_cluster_cat} available for this galaxy to identify these compact clusters. These clusters are indicated by filled magenta circles in the left panel of Fig.~\ref{fig:bubble_SFH}. Such clusters are also present in some of the small bubbles that are not part of our big bubble sample. In several cases, PHANGS clusters\footnote{We include all PHANGS clusters, irrespective of their assigned class.} spatially coincide with sources classified as stars in our JWST/NIRCam photometric catalogue. 
Given that the PHANGS cluster catalogue is based on HST imaging, it is expected that some of the sources may have been misclassified as extended sources due to crowding and the comparatively lower spatial resolution of HST, which are resolved into multiple sources in the JWST images.

In Fig.~\ref{fig:clusters}, we show all the PHANGS-HST clusters seen inside the bubbles in the color-color diagram (top) and CMD (bottom) using the photometry given in the PHANGS-HST catalogue. The diagram includes the SSP models from \citet{bc03}~(BC03) for a cluster of solar metallicity with no reddening corrections. The majority of the clusters have ages between 1--30~Myr, with the other clusters also consistent with ages less than $t_0$ with moderate amounts of reddening ($A_V<$2~mag) and cluster mass in the range 10$^3$--10$^4$~M$_{\sun}$. These cluster masses are systematically less than the accumulated stellar mass in each bubble. Besides, these clusters in Fig.~\ref{fig:bubble_SFH} are not preferentially located at the geometrical centres of the bubbles, indicating that they are unlikely to be precursor clusters responsible for triggering bubble formation. Instead, they are most likely formed in the dense parts of the expanding shells. To assess this quantitatively for the big-bubble sample, we examine the spatial and temporal relationship between clusters and their host bubbles by plotting the ratio of the cluster age ($t_c$)\footnote{$t_c$ was obtained by comparing SSP colors to observed cluster colors shown in Fig.~\ref{fig:clusters}} to the bubble age ($t_{25}$) as a function of the cluster-centric distance normalized by the bubble's effective radius, $r/R_{\mathrm B}$ (Fig.~\ref{fig:tc}). Genuine precursor cluster candidates are expected to lie near the bubble centre ($r/R_{\mathrm B} \approx 0$) and to have ages comparable to the bubble age ($t_c/t_{25} \approx 1$). 
Instead, we find that the majority of clusters projected within the bubbles are significantly younger than $t_{25}$. In the few cases where cluster ages are comparable to $t_{25}$, the clusters are preferentially located near the bubble rims. This spatial–temporal distribution strongly indicates that these cluster candidates are unlikely to be precursor clusters responsible for initiating bubble expansion. Rather, they are more consistent with having formed in the dense, swept-up material of the expanding shells. However, we note that there could be some clusters below the detection limit of the HST cluster catalog. Based on the V band detection limit (m$_V$=23.0) for NGC\,628 \citep{phangs_cluster_cat}, the cluster catalog is only sensitive to clusters with mass more than 6000~M$_{\sun}$, for clusters as old as 50~Myr.

\subsection{Stellar populations responsible for the creation of small bubbles}

We have 74 bubbles from our main sample where the JWST/NIRCam CMDs did not have a sufficient number of stars (10 stars younger than 50~Myr) for the construction of the SFH. In Fig.~\ref{fig:bubble_SFH_small}, we show the colour-composite images of four examples of these smaller bubbles (left) along their CMDs (middle) and age histograms (right). These being candidates for low-mass clusters or OB associations, they are not expected to contain a continuous distribution of stellar masses along the isochrones, due to the stochastic nature that gives rise to an IMF. Nevertheless, the few young (age$<$50~Myr) stars that are present in these bubbles cannot be fit by single isochrones, suggesting a large age spread of the stellar populations within these bubbles. This suggests that these small bubbles are created by one or a few massive stars, with newer stars forming within the bubble later on, most likely in the dense parts of the expanding shell.

Three of the 74 small bubbles contain a PHANGS-HST cluster candidate inside the bubble, located close to the geometrical centre of the bubble. In four cases, we identify isolated resolved stars near the bubble centres, while the corresponding IMF-associated stellar population is absent from the CMD. These sources could be partially resolved cluster candidates or OB associations.

Out of the 74 small bubbles in our sample, 51 have at least one star younger than 50~Myr and are also plotted in the fundamental plane for feedback-driven bubbles in Fig.~\ref{fig:energy}. They also follow the fundamental relation, occupying the low size end of the relation. The figure shows a tendency for these small bubbles to lie in the high $n/\eta$ regime. However, we cannot ascertain this tendency given that the stellar masses responsible for the creation of these bubbles could be less than the derived age due to the low number of stars detected in our CMD, which could bring them to a lower $n/\eta$ regime.

We now briefly discuss the origin of the subsample of 23 bubbles with radii predominantly less than 50~pc that show no evidence for either young (age $<$ 50~Myr) stars in the resolved population or a partially resolved PHANGS-HST cluster candidate. Three examples of this class of bubbles are shown in Fig.~\ref{fig:bubble_SFH_vsmall}. Starless bubbles are expected by exotic events such as a gamma-ray burst \citep{grb} or a high velocity cloud \citep{hvc} impinging on the disk. However, such events are expected to create bubbles of radius greater than 100~pc. Given their relatively small sizes, these starless bubbles most likely represent single-star events where the responsible massive star has exploded as a supernova, with no secondary star formation. Alternatively, the star responsible for creating these bubbles may be a runaway star, which currently lies outside the bubble. Incidentally, there are young (age $<$ 50~Myr) stars in the periphery of most of these starless bubbles, which could be these runaway stars. However, most of the starless bubbles are found in active star-forming sites, which makes an unambiguous association of the likely runaway stars with the nearest starless bubble a difficult task. We also note that some of these bubbles are hosted in crowded regions, hence insensitive to our star and cluster catalog. Future elaborative studies incorporating multi-wavelength data would be required to elucidate the nature of these interesting objects.

\subsection{The HI-hole}
The HI hole is the largest bubble in our sample (Bubble \#0) and is nearly twice the size of the next largest structure, the Phantom Void. Despite its large size, the HI hole broadly follows the same scaling relations between bubble radius, enclosed stellar mass, and stellar population age as the smaller bubbles in the sample. The only notable deviation appears in the velocity versus $t_{25}$ relation, where the HI hole lies as an outlier. Given its large radius, the HI hole has likely broken out of the disk, and its expansion is therefore no longer in the energy-conservation regime modeled in the previous section. The SFH and the CMD both indicate continuous star formation over the past $\sim$100~Myr, which appears longer than the expected lifetime of a feedback-driven bubble in the shearing environment of a rotating galactic disk, unlike the H\,\textsc{i} holes in dwarf irregular galaxies studied in similar works \citep[e.g.][]{bubble_hst_cmd, bubble_hst_cmd2}. To assess whether the H\,\textsc{i} hole in NGC\,628 could survive the effects of rotational shear, we estimate the shear timescale, which is given by
\begin{equation}
    \tau_{\rm shear} \sim \left|R_{\rm b}\frac{d\Omega}{dR}\right|^{-1},
\end{equation}
where $\Omega$ is the angular frequency and $R_{\rm b}$ is the bubble radius. We adopt $\left|d\Omega/dR\right| = 9.87~[{\rm km\,s^{-1}\,kpc^{-2}}]$ at a galactocentric radius of 6~kpc based on the kinematic analysis of \citet{ngc628_kinematics}. This yields a shear timescale of $\tau_{\rm shear} \approx 120~{\rm Myr}$. These estimates suggest that the H\,\textsc{i} hole could survive the effects of rotational shear, although, based on the  H\,\textsc{i} hole age derived in this work, after $\sim100~{\rm Myr}$ the structure would likely become significantly elongated or fragmented due to differential rotation.

In the scenario where the HI hole is produced by a high-velocity cloud or a sub-halo impinging on the disk \citep{subhalo}, the oldest stellar population associated with the HI hole based on SFH analysis would not represent the age of the bubble, since the triggering energy would instead originate from the collisional wave produced by the perturbing high-velocity cloud or sub-halo. 

Finally, based on all the analyses presented in this work, we do not find any discernible difference between the origin of the HI hole \citep[e.g.,][]{hi_hole2} and that of superbubbles based on the resolved stellar populations projected within them. This is because most of the energy that powers the bubbles comes from secondary star formation. Whether the first trigger occurs due to the impinging high-velocity cloud, subhalo, or stellar feedback from a cluster, the subsequent evolution of the bubble seems to be similar.

\section{Conclusions}
In this study, we used JWST/NIRCam photometry of NGC\,628 from the JWST--FEAST program to determine the ages and spatial distribution of resolved stellar populations associated with JWST/MIRI bubbles. By reconstructing star formation histories (SFHs) within a Bayesian framework applied to the JWST/NIRCam F115W, F150W, and F200W bands, in combination with PARSEC+COLIBRI isochrones, we characterized the temporal and spatial evolution of stars projected inside these feedback-driven structures. For the larger bubbles in our sample, the CMD and SFH analyses reveal multiple stellar populations within the bubble interiors, indicating that successive generations of stars contribute to driving their expansion. In most cases, the precursor cluster mass is only a small fraction of the total stellar mass currently found inside the bubble, suggesting that star formation becomes progressively more efficient as the bubble evolves. Phantom Void is an exception to this general scenario, which formed most of its stellar mass at early epochs. The geometrical centre of the bubbles does not contain any partially resolved or unresolved object that could have been the surviving precursor cluster. On the other hand, the resolved population contains stars of multiple ages. This behaviour is consistent with an initial triggering by a low-mass cluster or an OB association, followed by their gradual dissolution as the bubble expands, giving rise to the more spatially dispersed stellar populations observed in mature systems.

The spatial distribution of stellar ages exhibits a radial gradient: younger stars are preferentially located near the bubble rims, while older stars are distributed throughout the interiors. This pattern supports a scenario in which star formation propagates outward, with new stars forming in the swept-up shell and subsequently contributing additional feedback that drives further expansion.

A subset of smaller bubbles shows no evidence for young stars or clusters either within their interiors or along their rims. These cases likely represent episodes where shell-driven star formation was inefficient or absent, possibly reflecting episodic triggering that depends on the physical conditions of the accumulated material in the shells.

We do not find a clear relation between the bubble size and the age of the precursor stellar population, which is due to the non-negligible role of other parameters, such as the mass of the stellar populations, gas density, feedback efficiency, etc., in affecting the size of bubbles. All observed bubbles, including the small bubbles, follow a fundamental plane relation, which we defined based on the theory of stellar feedback-driven bubbles in a uniform medium. 

Overall, these results support a picture in which stellar feedback both sculpts the surrounding interstellar medium and promotes sequential star formation along expanding shells. The systematically rising SFHs imply that star formation within bubbles is cumulative and self-propagating, with the most recent activity concentrated along the bright PAH-emitting rims. The JWST/MIRI bubbles in NGC\,628 therefore provide direct, spatially resolved evidence that stellar feedback can simultaneously regulate and stimulate star formation in normal disk galaxies. More broadly, this work highlights JWST's ability to resolve the interplay between stellar feedback and star formation on sub-bubble scales, offering new insight into how feedback-driven structures seed subsequent generations of stars.

\section*{Acknowledgements}

We thank the anonymous referee for the insightful comments that have improved the manuscript. ACk expresses gratitude to the Secretaría de Ciencia, Humanidades, Tecnología e Innovación (SECIHTI) for funding his PhD through a grant. AB acknowledges support from PRIN2022 (2022NEXMP8) for the project "Radiative Opacities For Astrophysical Applications". We also thank Dr. Ivanio Puerari, Dr. Raul Naranjo, and Dr. Manuel Zamora for providing access to the OLINKI and Mixli clusters at INAOE, which were crucial for the computational work carried out in this research.

Computational resources were also provided by El Laboratorio Nacional de Supercómputo del Sureste de México (LNS), under project numbers \textsf{202404072C} and \textsf{202404078C}.

Further, this work incorporates observations from the PHANGS-MUSE large program \citep{phangs_muse} and data products from ESO observations under the following programs: 1100.B-0651, 095.C-0473, and 094.C-0623 (PHANGS-MUSE; PI: Schinnerer); 094.B-0321 (MAGNUM; PI: Marconi); 099.B-0242, 0100.B-0116, 098.B-0551 (MAD; PI: Carollo); and 097.B-0640 (TIMER; PI: Gadotti). We acknowledge the ESO Science Archive Facility for enabling access to these resources.

\section{Supplementary Material}

Supplementary material is available online and consists of:
\begin{itemize}
    \item Table S1 and Table S2, containing the full versions of Table~\ref{tab:bubble_sample} and Table~\ref{tab:bubble_properties}, respectively.
    \item Figures S3--S5, showing the complete bubble sample in a format analogous to Fig.~\ref{fig:bubble_SFH}, Fig.~\ref{fig:bubble_SFH_small}, and Fig.~\ref{fig:bubble_SFH_vsmall}, respectively.
\end{itemize}

%%%%%%%%%%%%%%%%%%%%%%%%%%%%%%%%%%%%%%%%%%%%%%%%%%
\section*{Data Availability}
This work is based on data obtained with the NASA/ESA/CSA James Webb Space Telescope, archived at the Mikulski Archive for Space Telescopes at the Space Telescope Science Institute. The institute is managed by the Association of Universities for Research in Astronomy, Inc., under NASA contract NAS 5-03127 for JWST. These data were collected as part of program \#1783 (JWST-FEAST; PI: Adamo Angela) and can be accessed at doi:\href{http://dx.doi.org/10.17909/1zxh-fx32}{10.17909/1zxh-fx32}.\\ \\

\textit{Facilities}: JWST/NIRCam and VLT/MUSE.

\textit{Softwares}: \href{http://americano.dolphinsim.com/dolphot/nircam.html}{DOLPHOT} \citep{dolphot_1, dolphot_2}, Astropy \citep{astropy}, Numpy \citep{numpy}, and Matplotlib \citep{matplotlib}



\bibliographystyle{mnras}
\bibliography{example} % if your bibtex file is called example.bib

%\begin{thebibliography}{99}
%\end{thebibliography}

%%%%%%%%%%%%%%%%%%%%%%%%%%%%%%%%%%%%%%%%%%%%%%%%%%

%%%%%%%%%%%%%%%%% APPENDICES %%%%%%%%%%%%%%%%%%%%%
\appendix
\section{Disk SFH grids}

\begin{figure*}
    \centering
    \includegraphics[width=0.8\linewidth]{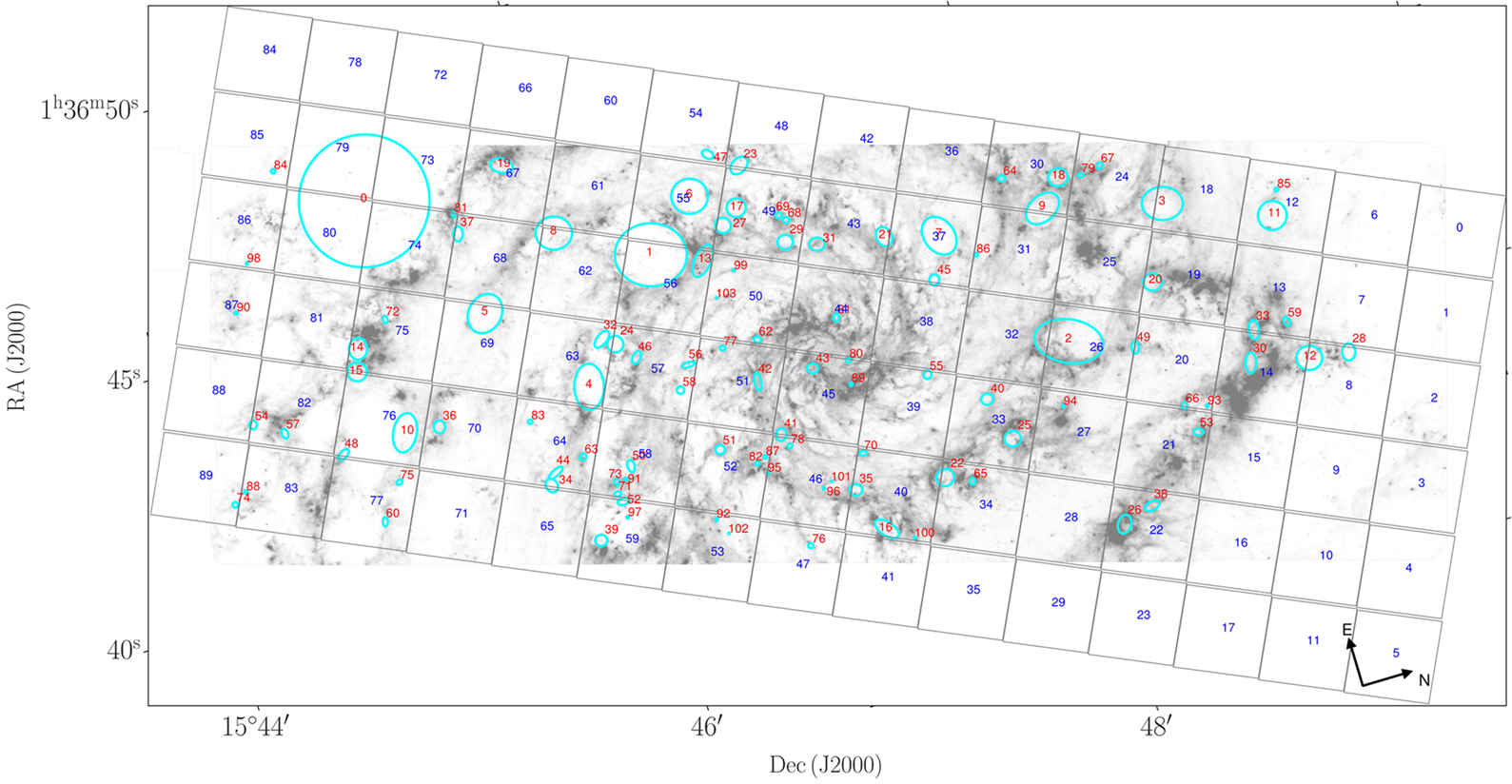}
    \caption{The grey background indicates the JWST/MIRI F770W image. The cyan ellipses indicate the bubbles studied in this work. The grey grid overlay corresponds to one utilized for estimating disk SFH.}
    \label{fig:grids}
\end{figure*}

Fig~\ref{fig:grids} is auxiliary to Fig~\ref{fig:bubble_sample}, where we show the bubbles overplotted on the F770W image along with the $\sim$kpc$^2$ grids used for estimating quantities such as completeness, MUSE gas-phase metallicity, TRGB-based metallicity, and extinction measurements and disk SFH.

\section{Bayesian Star Formation History}
\label{sec:bay_model}

\textbf{Bayes' theorem in hierarchical form:}
\begin{equation}
    p(a|F)\propto p(a)\prod_{j=1}^{N_D}\prod_{k=1}^{N_F}\int\dfrac{S(F_j^k)L(F_j^k|\hat F_j^k)p(f_j^k|a)}{l(a,S)}df_j^k
\end{equation}

\noindent
\textbf{Likelihood:}
\begin{equation}
    L(F_j^k|\hat F_j^k) = P(F_j^k, e_j^k|\hat F_j^k,\hat e_j^k) = (2\pi \bar e_j^k)^{-1/2}.\exp{\left (\frac{F_j^k-\hat F_j^k}{\bar e_j^k}\right )^2}
\end{equation}

\noindent
\textbf{Prior:}
\begin{equation}
    p(f_j^k|a) = \sum_{i=1}^{N_{ISO}}a_i W_i\int_{M_{l,i}}^{M_{u,i}}\Phi(M)\prod_{k=1}^{N_F}\mathcal{N}(F_j^k, e_j^k|\hat F_j^k,\hat e_j^k)dM
\end{equation}

\noindent
\textbf{Normalization:}
\begin{equation}
    l(a|S) = \sum_{i=1}^{N_{ISO}}\int_{M_{l,i}}^{M_{u,i}}\Phi(M)S(F_j^k) dM
\end{equation}

Where,

$N_D$ is the number of stars in the CMD.

$N_F$ is the number of photometric bands.

$N_{ISO}$ is the number of isochrones.

$
  \Phi(M) =
  \begin{cases}
    C_1M^{-\alpha_1} & M<M_c\\
     C_2M^{-\alpha_2} & M_c<M<M_u\\
  \end{cases}
$

$\bar e_j^k = \sqrt{(e_j^k)^2 + (\hat e_j^k)^2 }$

$W_i= \dfrac{\Delta t_i}{\sum_{i=1}^{N_{ISO}} \Delta t_i}$

\section{Subjacent Disk SFH}
\label{sec:sub_disk}
\begin{figure*}
    \centering
    \includegraphics[width=0.8\linewidth]{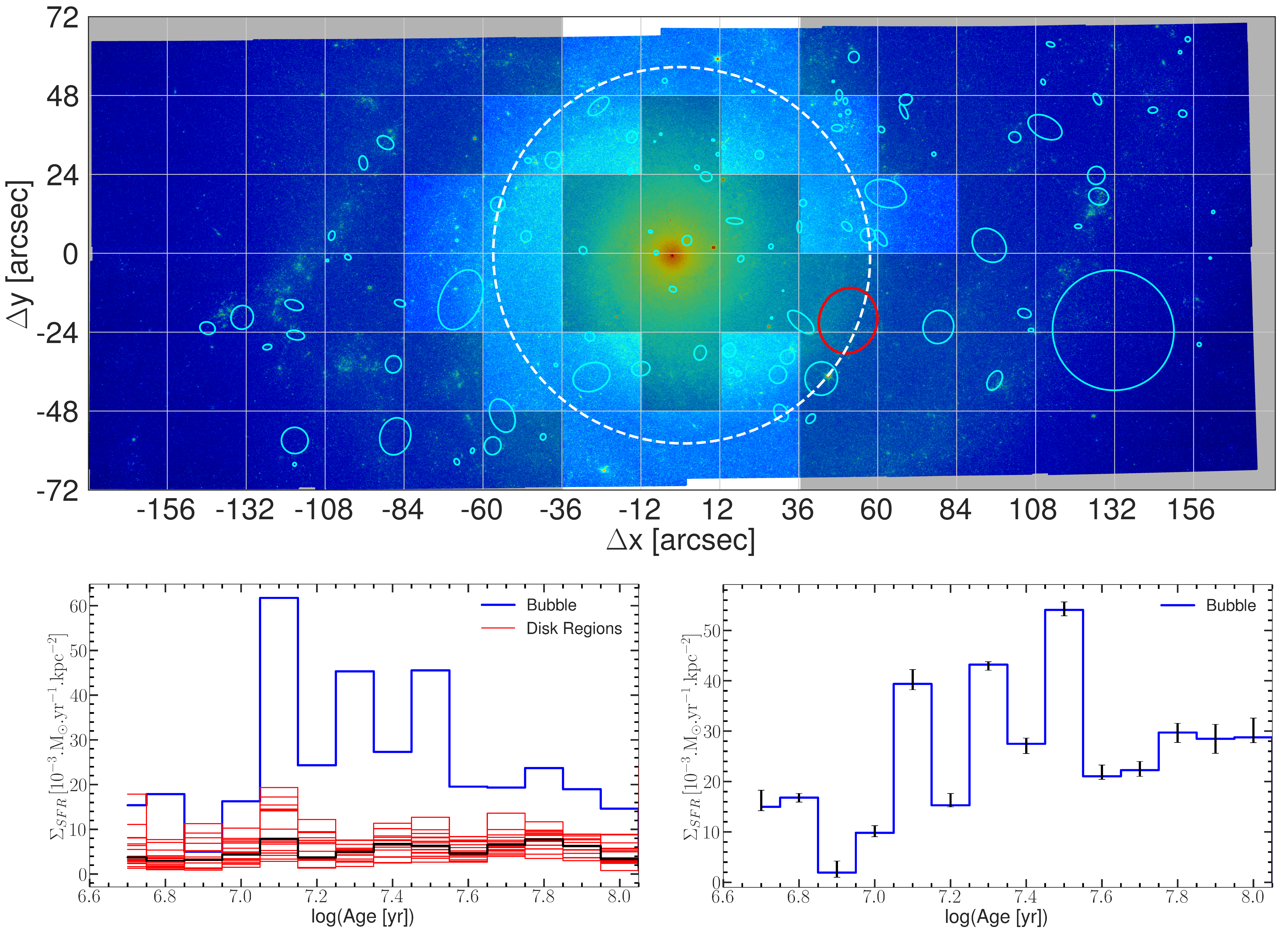}
    \caption{Illustration of the procedure adopted to obtain pure SFR of the stars involved in the bubble formation by subtracting the SFH of the adjacent disk from the observed SFH. (\textit{Top:}) Image in F115W showing all the bubbles (cyan ellipses) and the areas (highlighted grids) used to determine typical disk SFH. The bubble under analysis is shown by a red ellipse. The median of the SFHs of all grids that are at the same galactocentric distance as the bubble (white dashed circle) is used as the typical SFH of the stars from the subjacent disk seen projected inside the bubble. (\textit{Bottom Left:}) The blue step plot shows the $\Sigma_{\rm SFR}$ of the bubble. The red step plots show the $\Sigma_{\rm SFR}$ of different subjacent disk regions, with the black step plot showing the median SFH of the subjacent disk. \textit{(Bottom Right:)} The blue step plot shows the disk-subtracted $\Sigma_{\rm SFR}$ of the bubble, corrected for incompleteness (see Fig.~\ref{fig:bubble_SFH} for the rest of the details in the figure).}
    \label{fig:sub_disk}
\end{figure*}

Stars seen projected inside the bubble are the sum of the stars from the subjacent disk population plus any stars associated with the event that created the bubble. We are interested in isolating the latter, which can be obtained by subtracting the disk population at that region. The latter can be estimated by using the typical stellar populations at the same galactocentric radius as the bubble.

We use the azimuthal regions at the same galactocentric radius to obtain the typical stellar population, rather than regions adjacent to the bubble, because often the adjacent regions are star-forming in the spiral arms. Since there is no ongoing star formation inside the bubble, using the adjacent disk region results in oversubtraction of the young stellar populations. Obtaining the median of the azimuthal regions ensures that we are approximating a non-star-forming region at that galactocentric radius.

An example of the grids used for azimuthal median averaging for estimating the subjacent disk of an example bubble (Bubble \#1) is shown in Appendix Fig.~\ref{fig:sub_disk}. In the figure (top panel), all grid regions that are at the same galactocentric radius as the example are highlighted, except the region(s) containing the bubble. SFH for each of these grids is shown in red in the middle panel, with their median indicated by the black line. The bottom panel shows the subjacent disk-subtracted SFH of the bubble. As described in Section~\ref{sec:incomp}, the bubble SFH is corrected for incompleteness as well.

\section{SFR from MUSE H$\alpha$}
To obtain the H${\alpha}$ SFR within the bubble and shell using MUSE data, we applied several filters to ensure that only star-forming spaxels are selected.
Specifically, we required that:

\begin{itemize}
    \item $\mathrm{S/N} \geq 3$ for the H${\alpha}$, H${\beta}$,
    [O\,\textsc{iii}]$\lambda$5007, [N\,\textsc{ii}]$\lambda$6583, and [S\,\textsc{ii}]$\lambda$6717+6731 emission lines.
    
    \item Spaxels satisfy the star-forming criteria in both the
    [O\,\textsc{iii}]/H${\beta}$ vs.\ [N\,\textsc{ii}]/H${\alpha}$ and
    [O\,\textsc{iii}]/H${\beta}$ vs.\ [S\,\textsc{ii}]/H${\alpha}$ BPT
    diagrams \citep{bpt,bpt2,bpt3}.
    
    \item EW(H${\alpha}) \geq 50$\,\AA, which corresponds to a stellar population younger than 5~Myr \citep{ha_ew}, where
    EW(H${\alpha}$) is the H${\alpha}$ emission equivalent width.
\end{itemize}

The SFR is calculated using the \citet{ha_sfr} relation using calibrations from \citet{ha_sfr2}.
\begin{equation}
    {\rm SFR} [{\rm M}_{\sun}\,{\rm yr}^{-1}] = 5.4\times10^{-42}L(H\alpha), 
\end{equation}
where  L(H$\alpha$) is the extinction corrected H$\alpha$ luminosity in erg\,s$^{-1}$.

% Don't change these lines
\bsp	% typesetting comment
\label{lastpage}
\end{document}